\documentclass{aastex701}

\usepackage{hyperref}

\usepackage{enumitem,amssymb}
\newlist{todolist}{itemize}{2}
\setlist[todolist]{label=$\square$}
\usepackage{pifont}

\newcommand{\um}{$\mu$m}
\newcommand{\spherex}{SPHEREx}
\newcommand{\skysim}{Simulator}
\newcommand{\nw}{nW m$^{-2}$ sr$^{-1}$}
\newcommand{\MJysr}{MJy sr$^{-1}$}
\newcommand{\eps}{e$^{-}$s$^{-1}$}
\newcommand{\affilcaltech}{\affiliation{Department of Physics, California Institute of Technology, Pasadena, CA 91125, USA}}
\newcommand{\affiljpl}{\affiliation{Jet Propulsion Laboratory, California Institute of Technology, 4800 Oak Grove Drive, Pasadena, CA 91009, USA}}
\newcommand{\affilkasi}{\affiliation{Korea Astronomy and Space Science Institute (KASI), 776 Daedeok-daero, Yuseong-gu, Daejeon 34055, Republic of Korea}}
\newcommand{\affilipac}{\affiliation{IPAC, California Insitute of Technology, 770 S. Wilson Ave, Pasadena, CA 91125, USA}}
\newcommand{\secref}[1]{Sec.~\ref{#1}}
\newcommand{\figref}[1]{Fig.~\ref{#1}}

\newdimen\sa  \newdimen\sb
\def\arcsec{\ifmmode {^{\scriptstyle\prime\prime}}
          \else $^{\scriptstyle\prime\prime}$\fi}
          
\def\arcmin{\ifmmode {^{\scriptstyle\prime}}
          \else $^{\scriptstyle\prime}$\fi}     
          
\def\parcs{\sa=.07em \sb=.03em
     \ifmmode \hbox{\rlap{.}}^{\scriptstyle\prime\kern -\sb\prime}\hbox{\kern -\sa}
     \else \rlap{.}$^{\scriptstyle\prime\kern -\sb\prime}$\kern -\sa\fi}

\def\deg{\ifmmode^\circ\else$^\circ$\fi}

\begin{document}
\submitjournal{ApJS}
\makebox[17cm][r]{\textcopyright 2025 all rights reserved.}

\title{The \spherex\ Sky Simulator: Science Data Modeling for the First All-Sky Near-Infrared Spectral Survey}

\author[0000-0002-4650-8518]{Brendan P. Crill}
\email{bcrill@jpl.nasa.gov}
\affiljpl


\author[0000-0002-2618-1124]{Yoonsoo P. Bach}
\email{ysbach@kasi.re.kr}
\affilkasi

\author[0000-0003-4607-9562]{Sean A. Bryan}
\email{sean.a.bryan@asu.edu}
\affiliation{School of Earth and Space Exploration, Arizona State University, Tempe, AZ 85287 USA}

\author[0009-0008-8066-446X]{Jean Choppin de Janvry}
\email{jeanchdj@lbl.gov}
\affiljpl
\affiliation{University of California at Berkeley, Berkeley, CA 94720, USA}
\affiliation{Lawrence Berkeley National Laboratory, 1 Cyclotron Road, Berkeley, CA 94720, USA}

\author[0000-0002-7471-719X]{Ari J. Cukierman}
\email{ajcukier@caltech.edu}
\affilcaltech

\author[0009-0002-0098-6183]{C.~Darren Dowell}
\email{charles.d.dowell@jpl.nasa.gov}
\affiljpl

\author[0000-0002-3745-2882]{Spencer W. Everett}
\email{severett@caltech.edu}
\affilcaltech

\author[0000-0001-9925-0146]{Candice Fazar}
\email{cmfsps@rit.edu}
\affiliation{School of Physics and Astronomy, Rochester Institute of Technology, Rochester, NY 14623, USA}

\author[0009-0003-5316-5562]{Tatiana Goldina}
\email{tatianag@ipac.caltech.edu}
\affilipac

\author[0009-0009-1219-5128]{Zhaoyu Huai}
\email{zhuai@caltech.edu}
\affilcaltech

\author[0000-0001-5812-1903]{Howard Hui}
\email{hhui@caltech.edu}
\affilcaltech

\author[0000-0002-2770-808X]{Woong-Seob Jeong}
\email{jeongws@kasi.re.kr}
\affilkasi

\author[0000-0002-3470-2954]{Jae Hwan Kang}
\email{jkang7@caltech.edu}
\affilcaltech

\author[0009-0003-8869-3651]{Phillip M. Korngut}
\email{phil.m.korngut@jpl.nasa.gov}
\affilcaltech

\author[0000-0003-0894-7824]{Jae Joon Lee}
\email{leejjoon@kasi.re.kr}
\affilkasi

\author[0000-0000-0000-0000]{Daniel C. Masters}
\email{dmasters@ipac.caltech.edu}
\affilipac

\author[0000-0001-9368-3186]{Chi H. Nguyen}
\affilcaltech
\email{chnguyen@caltech.edu}

\author[0000-0001-9937-8270]{Jeonghyun Pyo}
\email{jhpyo@kasi.re.kr}
\affilkasi

\author[0000-0002-9554-1082]{Teresa Symons}
\email{tsymons@ipac.caltech.edu}
\affilipac

\author[0000-0003-3078-2763]{Yujin Yang}
\email{yyang@kasi.re.kr}
\affilkasi

\author[0000-0001-8253-1451]{Michael Zemcov}
\email{mbzsps@rit.edu}
\affiliation{School of Physics and Astronomy, Rochester Institute of Technology, Rochester, NY 14623, USA}
\affiljpl


\author[0000-0001-9674-1564]{Rachel Akeson}
\email{rla@ipac.caltech.edu}
\affilipac

\author[0000-0002-3993-0745]{Matthew L.\ N.\ Ashby}
\email{mashby@cfa.harvard.edu}
\affiliation{Center for Astrophysics  $|$ Harvard \& Smithsonian,
Optical and Infrared Astronomy Division, Cambridge, MA 01238, USA}

\author{James J. Bock}
\email{jjb@astro.caltech.edu}
\affiljpl
\affilcaltech

\author[0000-0002-5437-0504]{Tzu-Ching Chang}
\email{tzu-ching.chang@jpl.nasa.gov}
\affiljpl

\author[0000-0002-5437-0504]{Yun-Ting Cheng}
\email{ycheng3@caltech.edu}
\affilcaltech

\author[0000-0001-6320-261X]{Yi-Kuan Chiang}
\email{ykchiang@asiaa.sinica.edu.tw}
\affiliation{Academia Sinica Institute of Astronomy and Astrophysics (ASIAA), No. 1, Section 4, Roosevelt Road, Taipei 10617, Taiwan}

\author[0000-0002-3892-0190]{Asantha Cooray}
\email{acooray@uci.edu}
\affiliation{Department of Physics \& Astronomy, University of California Irvine, Irvine CA 92697, USA}

\author[0000-0002-0867-2536]{Olivier Dor\'{e}}
\email{olivier.p.dore@jpl.nasa.gov}
\affiljpl
\affilcaltech

\author[0000-0002-9382-9832]{Andreas L. Faisst}
\email{afaisst@ipac.caltech.edu}
\affilipac

\author[0000-0002-9330-8738]{Richard M. Feder}
\email{rmfeder@berkeley.edu}
\affiliation{University of California at Berkeley, Berkeley, CA 94720, USA}
\affiliation{Lawrence Berkeley National Laboratory, 1 Cyclotron Road, Berkeley, CA 94720, USA}

\author[0000-0003-4990-189X]{Michael W. Werner}
\email{michael.w.werner@jpl.nasa.gov}
\affiljpl

\begin{abstract}
We describe the \spherex\ Sky Simulator, a software tool designed to model science data for NASA's \spherex\ mission that will carry out a series of all-sky spectrophotometric surveys at $\sim$6\arcsec spatial resolution in 102 spectral channels spanning 0.75 to 5 $\mu$m.  The \skysim\ software implements models for astrophysical emission, instrument characteristics, and survey strategy to generate realistic infrared sky scenes as they will be observed by \spherex.  The simulated data includes a variety of realistic noise and systematic effects that are estimated using up-to-date astrophysical measurements and information from pre-launch instrument characterization campaigns.  Through the pre-flight mission phases \skysim\ has been critical in predicting the impact of various effects on \spherex\ science performance, and has played an important role guiding the development of the \spherex\ data analysis pipeline.  In this paper, we describe the \skysim\ architecture, pre-flight instrument and sky models, and summarize high-level predictions from the \skysim\, including a pre-launch prediction for the 5$\sigma$ point source sensitivity of \spherex,  which we estimate to be $m_{\rm AB}$ 18.5--19 from 0.75 to 3.8~$\mu$m and $m_{\rm AB}$ 16.6--18 from 3.8 to 5 $\mu$m, with the sensitivity limited by the zodiacal light background at all wavelengths.  In the future, on-orbit data will be used to improve the \skysim\, which will form the basis of a variety of forward-modeling tools that will be used to model myriad instrumental and astrophysical processes to characterize their systematic effects on our final data products and analyses.  
\end{abstract}

\keywords{astronomical simulations -- astronomical instrumentation -- infrared spectroscopy -- sky surveys -- space telescopes}

\section{Introduction} \label{sec:intro}

\spherex\footnote{\url{http://spherex.caltech.edu}}, the Spectro-Photometer for the History of the Universe, Epoch of Reionization, and ices Explorer \citep{Dore2014}, is a NASA MIDEX mission launched 11 March, 2025. \spherex\ will perform the first all-sky near infrared spectral survey with spectral resolving power $\lambda / \Delta \lambda = R \sim 40$ between 0.75 and 3.8 \um\ and $R \sim 120$ between 3.8 and 5 \um. At the end of its two-year primary mission, \spherex\ will provide complete spectra of each 6\parcs{15} $\times$ 6\parcs{15} pixel on the sky.  This information will be used to perform the three primary science cases discussed in \citet{Crill2020} that include: mapping the 3-dimensional structure of the redshift $z \lesssim 2$ universe to constrain physics of inflation; performing intensity mapping to trace the development of star formation in galaxies over cosmic history; and tracing the history of water and biogenic molecules from the interstellar medium (ISM) into star systems.  In all, \spherex\ is expected to produce: $>10^{9}$ galaxy spectra; $> 10^{8}$ high-quality galaxy redshifts; $> 10^{8}$ stellar spectra; $> 10^{7}$ high-quality ice absorption spectra; $> 10^{6}$ quasar spectra; and $> 10^{4}$ asteroid spectra from objects in our own solar system.  The \spherex\ mission will provide images and tools to access the all-sky data set via NASA’s Infrared Science Archive (IRSA).  These will enable the global astrophysics community to investigate science questions spanning the entire field \citep{Dore2016, Dore2018}.  \spherex\ successfully completed in-orbit checkout on May 1, 2025 and began its 25-month survey on the same day.  

The \spherex\ mission and science team require science data simulations in all phases of the mission that are both astrophysically accurate and reflective of the properties of the instrument for a number of purposes, including: 1. supporting requirement definition during the design phases; permitting software architecture prototyping; 2. creating simulated data to assist in developing software modules; 3. validating pipeline software modules; and 4. predicting and analyzing the effects of systematic errors on the scientific results.  As the project progressed from the design and prototyping (mission Phases B \& C) through integration and testing (Phase D), past launch, to analyzing flight data at scale (Phase E), we expect to further develop the simulator to forward model various instrumental and astrophysical systematic errors to quantify their effect on the science products.

Because no available science data set can act as a proxy close enough to \spherex\ to sufficiently represent its unique instrument architecture, observation plan, and astrophysical constituents, the team turned to  simulation software.  A number of community-developed and -supported tools for simulating astrophysical signals (e.g. GalSim; \cite{Rowe2015}) and astronomical detectors (e.g. Pyxel; \cite{Prodhomme2020}) are available.  However, \spherex's unique spectrometer design with a custom on-bard detector readout, and the all-sky nature of the survey, led the \spherex\ team to develop a custom software tool called the \textit{\spherex\ Sky Simulator} (henceforth the \skysim).

In addition to providing a realistic representation of \spherex\ observations, we also imposed the requirements that the \skysim\ be  scalable to High Performance Computing (HPC) environments to enable large-scale simulations, portable to different HPC platforms; well-documented, and 
flexible enough to easily introduce additional systematic effects in the \spherex\ payload instrument as they are discovered and characterized.  

In this paper we describe the detailed implementation of the \skysim, including its overall architecture, the instrumental, observational and astrophysical algorithms and models it implements, and checks performed to validate its performance using characterization data acquired in the laboratory during the mission's Phase C.  Here we report on the version of the \skysim\ as of the end of Phase D, which includes our best estimates of the instrument before flight data were acquired.  Aside from providing a description of the tool, this paper also describes our expectations for the nature and type of \spherex\ data, as well as final pre-flight sensitivity estimates to assist the community in developing their own science cases as the data become public.  In outline, Section~\ref{sec:architecture} describes the overall architecture of the software tool, and Section~\ref{sec:data} provides an overview of the \spherex\ data products.  Next we describe the detailed implementation of the model: Section~\ref{sec:inputs} describes the astrophysical sky models used to populate the simulated images, Section~\ref{sec:survey} explains the survey planning software that generates the sequence of observations for the \spherex\ survey, and Section~\ref{sec:instrument} presents the instrument model used in the \skysim.  We summarize high-level outputs and results in Section~\ref{sec:results} including pre-launch point source sensitivity estimates from the \skysim.  Finally, in Sect.~\ref{sec:future} we present plans to further develop the \skysim\ software during the mission and maintain it as an essential  tool for the science data analysis.

\section{Architecture and Execution}
\label{sec:architecture}

\begin{figure}[!ht]
\plotone{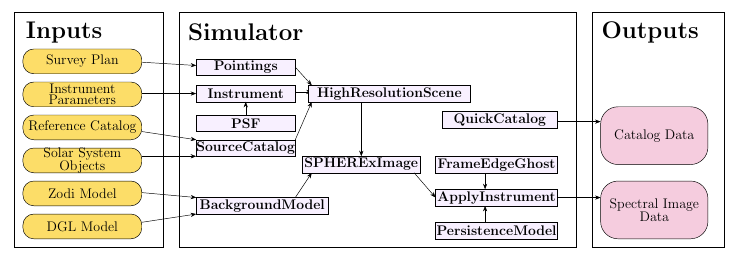}
\caption{\spherex\ \skysim\ architecture.}
\label{fig:architecture}
\end{figure}

To facilitate use on multiple platforms, the \skysim\ is coded entirely in Python and makes extensive use of community-developed and open-source packages such as NumPy \citep{harris2020} and Astropy \citep{astropy2022}.  The code is designed around a set of code classes that implement astrophysical sky signals, read a survey plan, model the instrument, and output data products (See Fig.~\ref{fig:architecture}).  The same \skysim\ code is used in both a personal computer environment for small scale simulations and in High Performance Computing (HPC) facilities for larger scale runs to simulate full-mission scale datasets.  

The instantiation of classes and execution order of functions is flexible and depends on the user's needs.  Most commonly, users want to generate simulated exposures, which requires calling functions that perform the following steps.  First, the parameters of the observations and the instrument are loaded into Pointings and Instrument classes.  The Pointings class is initialized by a survey plan (see Sect.~\ref{sec:survey}), which is a list of time-sequenced planned telescope pointings.  These are used to define the region of the sky to simulate for each exposure.  Next, instrument parameters are retrieved from a separate software repository that is shared with other major software elements including the Survey Planning Software (Sect.~\ref{sec:survey}) and the \spherex\ Science Data Center's reduction pipelines for consistency across the collaboration.  Following this, an astrophysical scene is simulated, which requires models for the various source populations.  A Reference Catalog (Sect.~\ref{sec:compactsources}) contains the list of compact sources to be injected into the simulation.  These are stored in files indexed by sky location, enabling only necessary portions of the sky to be loaded from disk.  Optionally, Solar System objects are injected by querying the JPL small body identification tool \citep{SBI}.  To speed up calculations, the \skysim\ has a shortcut `QuickCatalog' mode that can rapidly generate simulated photometry data for every planned observation of a selected subset of sky sources, as discussed in Sect.~\ref{sec:quickcatalog}.  This mode was used by \cite{Feder2024} to study the galaxy observations expected from \spherex, and also by \cite{Ivezic2021}, who explored opportunities for studying solar system objects using \spherex\ data.  It is also used extensively to simulate catalog data of thousands of ices sources and galaxies to test pipelines.  Finally, diffuse background models for the Zodiacal light (ZL) and diffuse galactic light (DGL) are loaded from disk and defined with user-selected options.

After setting up options and model definition, a loop over the list of pointings simulates each of the requested image simulations.  For each pointing and detector to be simulated, the transformation between sky coordinates and detector coordinates is determined based on the geometry of the telescope and the orientation of the spacecraft from the survey plan.    Using this rotation, compact sources are selected and the flux injected into a high resolution scene (HighResolutionScene class) as a delta function which is then convolved with a point spread function (PSF) model.  The scene is degraded to native \spherex\ resolution and diffuse signals are added.  The sample-up-the-ramp measurement of the  instrument (ApplyInstrument method) and other instrumental effects such as the frame edge ghost and image persistence are added. Finally, the result is written to a file that mimics the format as the \spherex\ flight data (Sect.~\ref{sec:data}).

Though users are typically interested in image simulations, the \skysim\ can produce outputs at different levels of processing with a wide range of optional inputs.  All of the models and their options discussed in this paper can be enabled (or disabled) by the user, with the only mandatory instantiations being the fundamental classes representing the survey, astrophysical emission, and instrument. 

\section{\spherex\ \skysim\ Data Products} 
\label{sec:data}

The \spherex\ data products are categorized into data levels (Table~\ref{tab:datalevels}), which correspond to different stages of processing, as is customary to many NASA missions.  Level 0 data are the packetized and compressed raw data from downlinks from the \spherex\ spacecraft.  Level 1 Image Data Files are the Level 0 data repacked into community-standard FITS data format, with each file corresponding to an image from a single detector for a single pointing of the \spherex\ Observatory, including assignment of coarse pointing information from the star tracking telescopes.  These data are minimally processed and include a variety of housekeeping information used in the data processing pipeline.  Level 2 data have been passed through a reduction pipeline that cleans the spectral image data of a variety of instrumental effects, associates pointing solutions, and calibrates from raw digital units to absolute intensity with units \MJysr\ (to be described in detail in future papers).  The Level 3 data comprise catalogs of compact source photometry used in the science analysis, and the Level 4 products are those used for science interpretation for the three science themes.  
The \skysim\ can produce Level 0, 1, and 2 data, but the typical uses for the outputs are either debugging and characterizing the Level 2 reduction pipeline, or to test and make predictions with downstream science analysis.  As a result, the \skysim\ data are typically output in the Level 2 format, which we describe in detail in Sect.~\ref{sS:spectral_images}, though Level 3 data can also be directly generated (Sect.~\ref{sec:quickcatalog}).

\begin{table}
    \centering
    \begin{tabular}{p{0.3in}llp{3in}} 
          \hline
         \hline
         Level&  Name& Units&Description\\ 
         \hline
         0&  Raw Image Data& e$^-$/s&Image data, including phantom pixel and extra reference pixel reads. \\ 
         1&  Engineering Image Data&  e$^-$/s&Image data corrected with phantom and reference pixel data.\\ 
         2&  Spectral Image Data&  MJy/sr&Calibrated spectral image data.\\ 
         3&  Catalog Data&  mJy&A source photometry catalog.\\ 
         \hline
         \hline
    \end{tabular}
    \caption{\spherex\ data products produced by the \skysim.  Levels 2 and 3 correspond to \spherex\ public data products.}
    \label{tab:datalevels}
\end{table}

The \skysim\ needs models of the sky and source catalogs to populate the images.  The generation of these catalogs is particularly subtle for \spherex\ as the spectral resolution is not similar to previous surveys, and all three science cases are critically dependent on the fidelity of the knowledge of sources to achieve their science.  We present the methods used to generate the \skysim\ catalogs in Section \ref{sS:catalogs}.

\subsection{Spectral Images}
\label{sS:spectral_images}

\begin{figure}
    \gridline{
        \fig{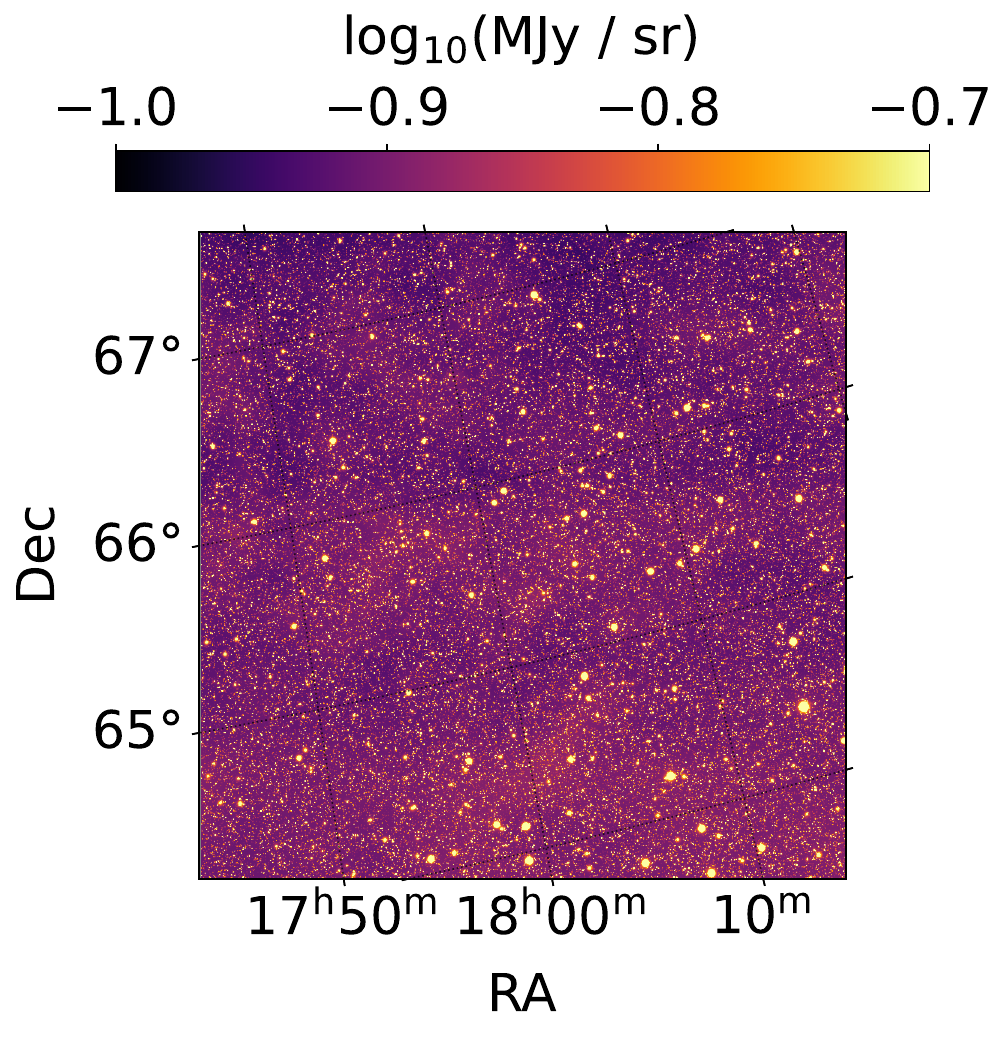}{0.33\textwidth}{Array~1}
        \fig{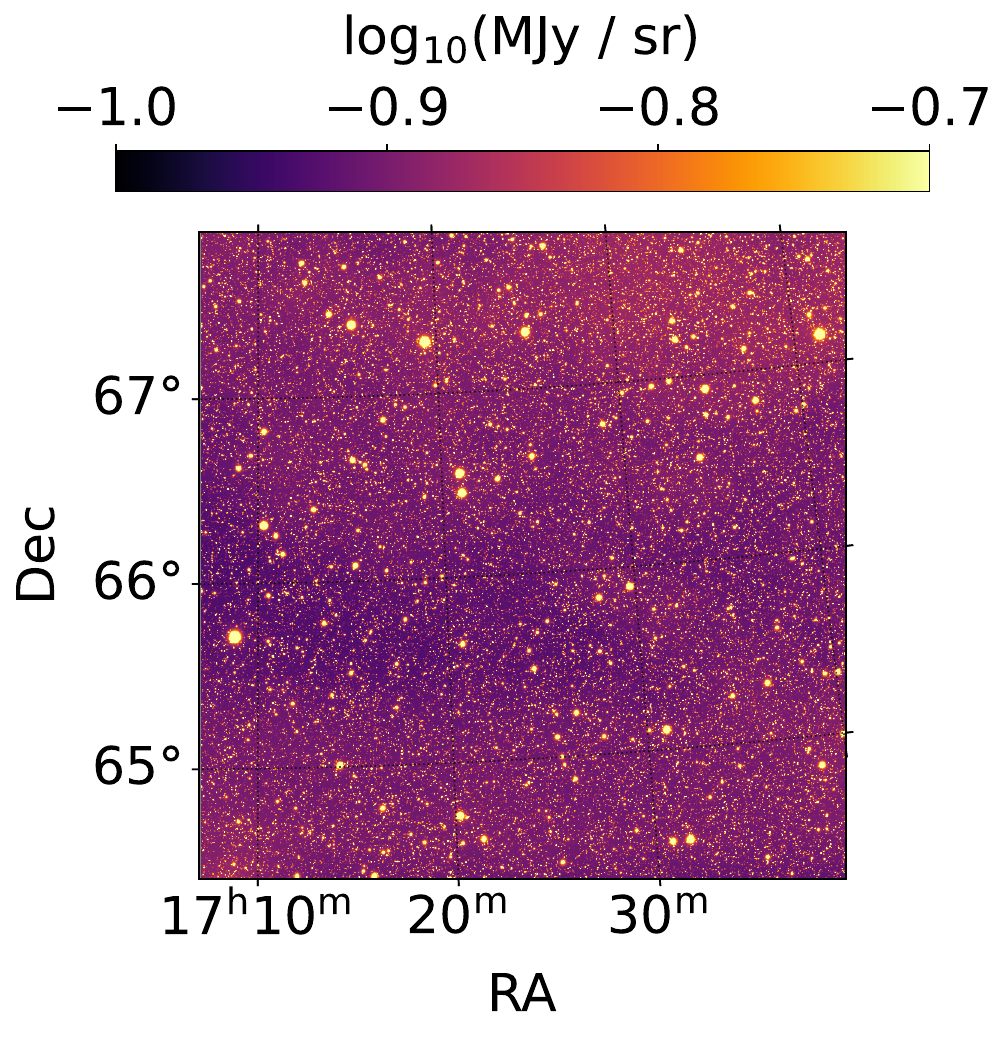}{0.33\textwidth}{Array~2}
        \fig{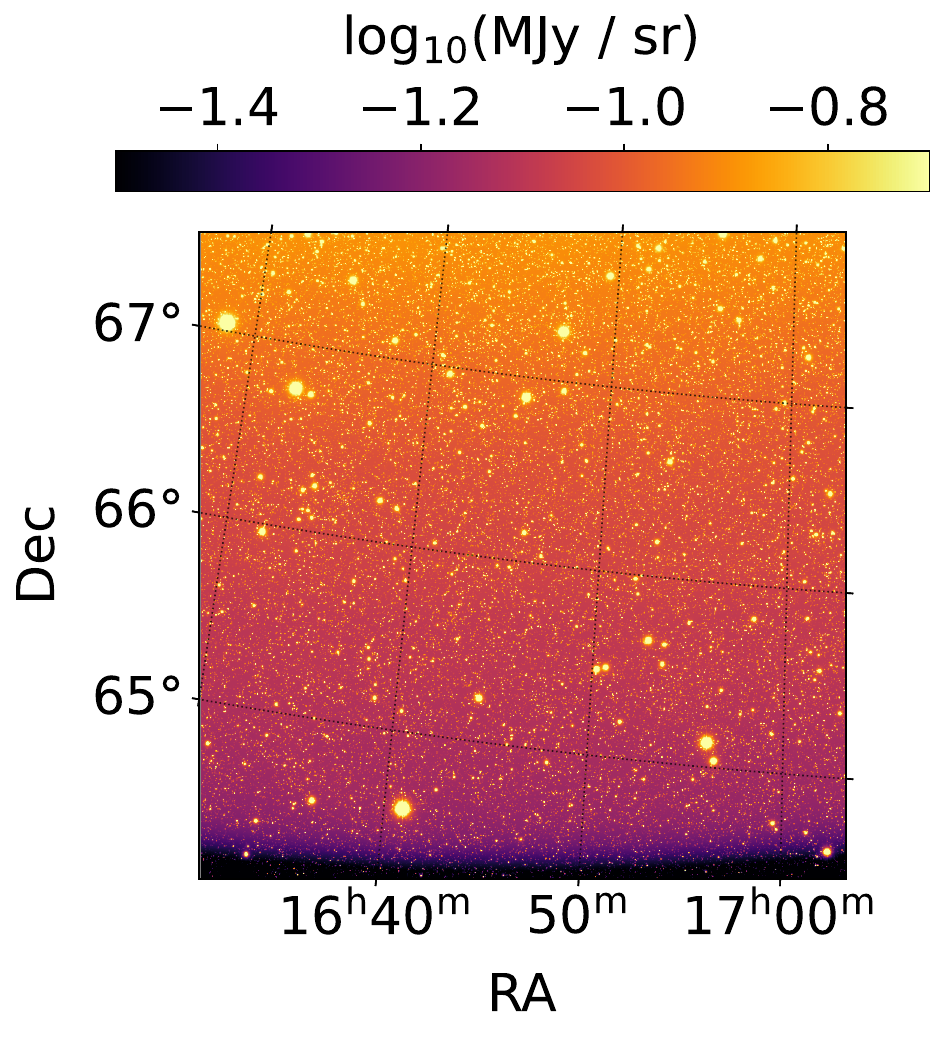}{0.3093596\textwidth}{Array~3}
    }
    \gridline{
        \fig{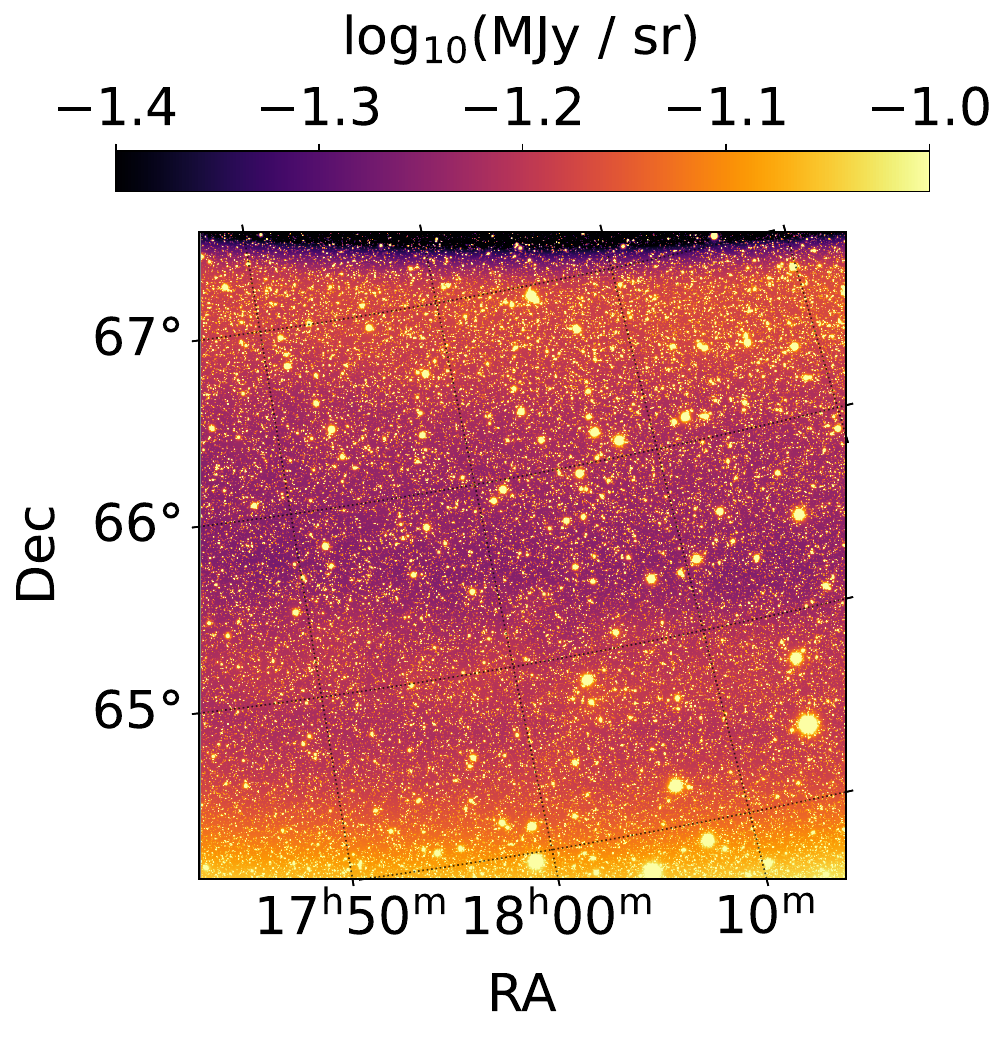}{0.33\textwidth}{Array~4}
        \fig{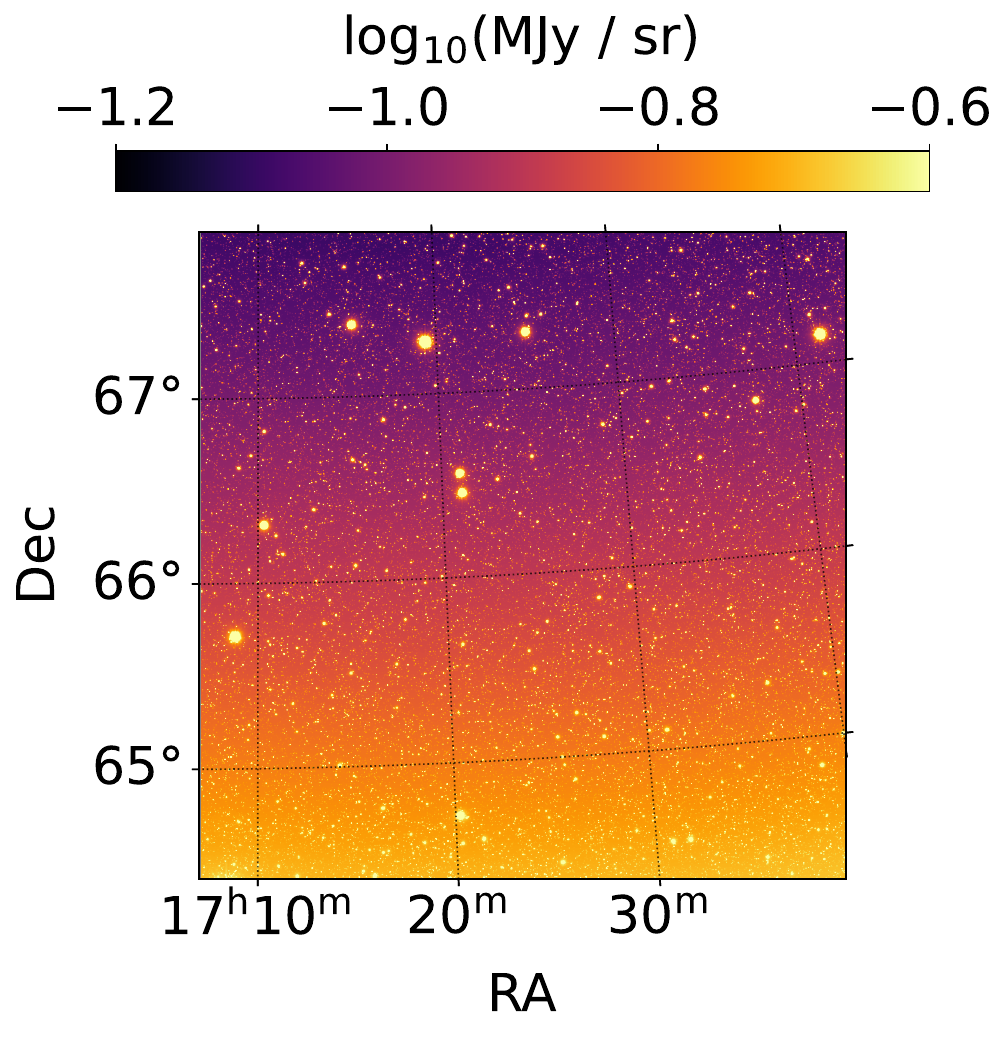}{0.33\textwidth}{Array~5}
        \fig{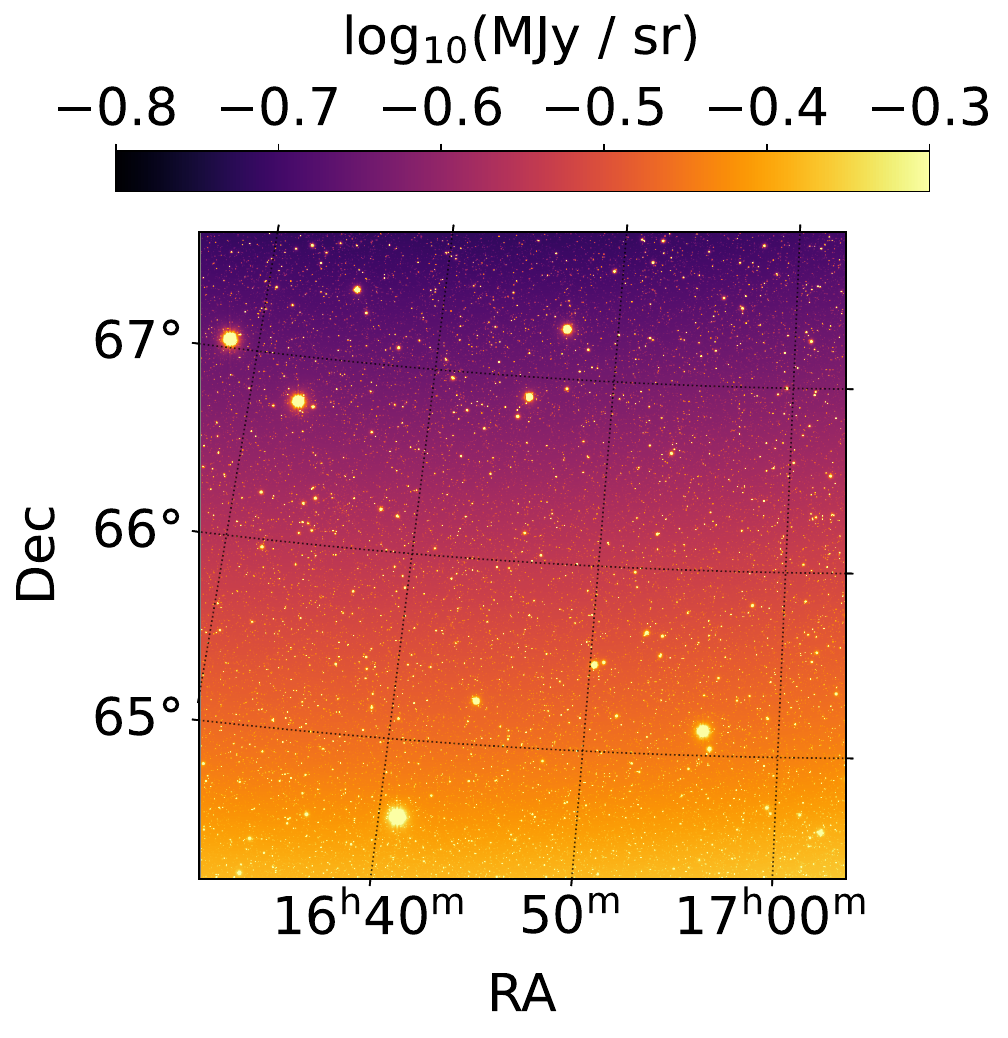}{0.33\textwidth}{Array~6}
    }
    \caption{Example simulated exposure images near the north ecliptic pole. In each image, wavelength increases in the downward direction (see Table~\ref{tab:spherex_bands}). The short-wavelength bands 1, 2, and 3 are in the top row, and the mid-wavelength bands 4, 5, and 6 are in the bottom row. The pointings are approximately equivalent for the two focal planes. The most prominent components are stars, Zodiacal light and diffuse Galactic light~(DGL). The Zodiacal light is spatially smooth with a steeply rising SED at long wavelengths (see Table~\ref{tab:spherex_bands} for wavelength coverage of each band); this creates a strong vertical gradient in the LVF~images for the longest wavelengths. The DGL manifests as nebular anisotropies that are visible in the shorter-wavelength arrays. Below the mid-point of the band-4 image, the horizontal band is due to a PAH emission feature at~$3.3~\mu\mathrm{m}$. At the bottom of band~3 and the top of band~4, there is a dimming due to the minimum in the optical efficiency of the dichroic beam splitter.}
    \label{fig:exampleImages}
\end{figure}

The \spherex\ spectral images are represented in the \skysim\ as $2048 \times 2048$ pixel images, one per detector, matching the layout of the six detectors in an exposure. Example simulated images are presented in \figref{fig:exampleImages}. The spectral axis is along one dimension of each detector, with roughly constant-wavelength regions along the other dimension.  Both axes are spatial images with a 6\parcs{15}/pixel plate scale.  Because the spectral bandpasss variation produced by an LVF is continuous, neighboring pixels in the spectral dimension are tuned to very similar wavelengths, with a slow gradient along that axis.  
To capture this behavior, we define 17 regions of equal area we call \textit{spectral channels} on each of the six detector arrays (making a total of 102 spectral channels).  The wavelength of peak response of all pixels within a spectral channel is within one spectral resolution element $\Delta\lambda$; the spectral channels are thus quasi-independent spectrally (Figs.~\ref{fig:SWIRbandpasses} and \ref{fig:MWIRbandpasses}).  The \spherex\ survey plan (Sect.~\ref{sec:survey}) is designed to place every source on the sky within each spectral channel at least once during a 6 month survey, ensuring a fully sampled spectrum.

The Level-2 images exclude the four-pixel-wide rim of pixels present on the edges of the H2RG detectors, resulting in 2040$\times$2040 science data arrays. The Level-2 data are stored in Flexible Image Transport System \citep{Pence2010} format by default, which is designed to match the public-facing data products.  A variety of meta-data describing the image processing and data provenance are included.  The FITS files also contain flag and mask information that mimics flight data.  For the \skysim\ only, a catalog of sources that have been populated into each images are optionally included to facilitate tests and analyses (see Section \ref{sS:catalogs}).  The \spherex\ team has made example Level 2 products available as Quick Look data through IRSA so the community can build tools to ingest sample data before flight data become public\footnote{Available at \url{https://irsa.ipac.caltech.edu/data/SPHEREx/AASJan2025/}. There may be discrepancies in the image headers between simualted and real exposures.}. 

\subsection{Catalog Data}
\label{sS:catalogs}

The \spherex\ science data pipeline will photometer known compact sources by carrying out forced photometry on the \spherex\ spectral image data at billions of pre-selected sky locations (Sect.~\ref{sec:compactsources}).  The \spherex\ photometry catalog serves as direct input into the Ices and Inflationary Cosmology science analyses, and for expediency, the \skysim\ can directly generate simulated photometry data without generating full spectral images (Sect.~\ref{sec:quickcatalog}).
The table of photometric measurements is designed to match the format of the \spherex\ catalog data.  The Primary Catalog consists of a table of photometry containing each separate measurement as a unique row.
After 12 months of photometry data becomes available, a Secondary Catalog is created by rebinning all of the photometric measurements onto a common wavelength grid (the spectral channels).  The Secondary Catalog also contains synthetic broadband photometry to match broadband measurements by other instruments, formed by averaging \spherex\ flux measurements weighted by bandpasses of 2MASS, WISE, and LSST.

\subsection{QuickCatalog mode}
\label{sec:quickcatalog}

\begin{figure}[ht!]
\plotone{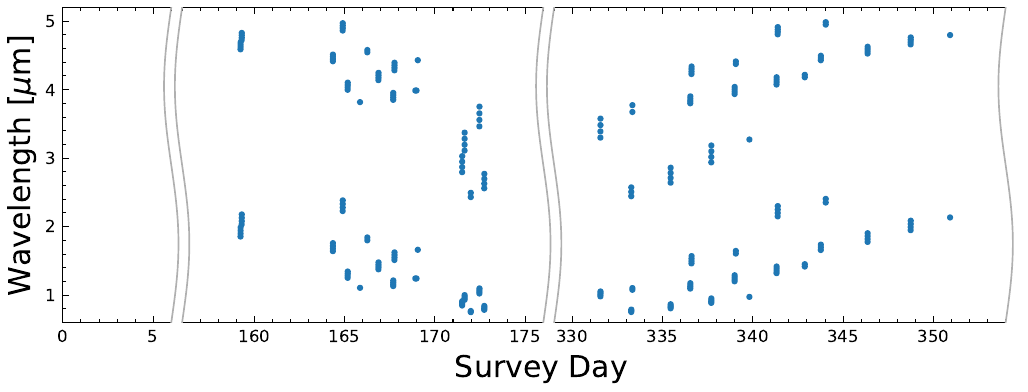}
\caption{Observation wavelengths vs. time (in days since the start of the mission) using QuickCatalog for a primary calibration source (GD71) in one mission year.  The survey plan covers this particular source in two distinct epochs in this year, and is representative of the coverage of most sources away from the ecliptic pole regions. \label{fig:quickcatalog_times}}
\end{figure}

A common simulation use case is to generate all of the \spherex\ photometry measurements during the mission for a single source or a subset of sources, bypassing the generation of full exposure data.  This is the \textit{QuickCatalog} mode of the \skysim.  
To use this feature, first a list of source coordinates is assembled, and the survey plan is used to find which \spherex\ pointings contain the sources in the field-of-view, and exactly where on the detector each source falls.  This uniquely defines the bandpass through which \spherex\ makes each individual observation.  An example of observation times and wavelengths for a typical low ecliptic latitude calibration source is shown in Fig.~\ref{fig:quickcatalog_times}.  A simulated subset of the detector array for each observations is generated by the \skysim, including backgrounds and the PSF, and with multiple realizations of noise sources to characterize noise properties.  This produces an image-space representation.

\begin{figure}[ht!]
\plotone{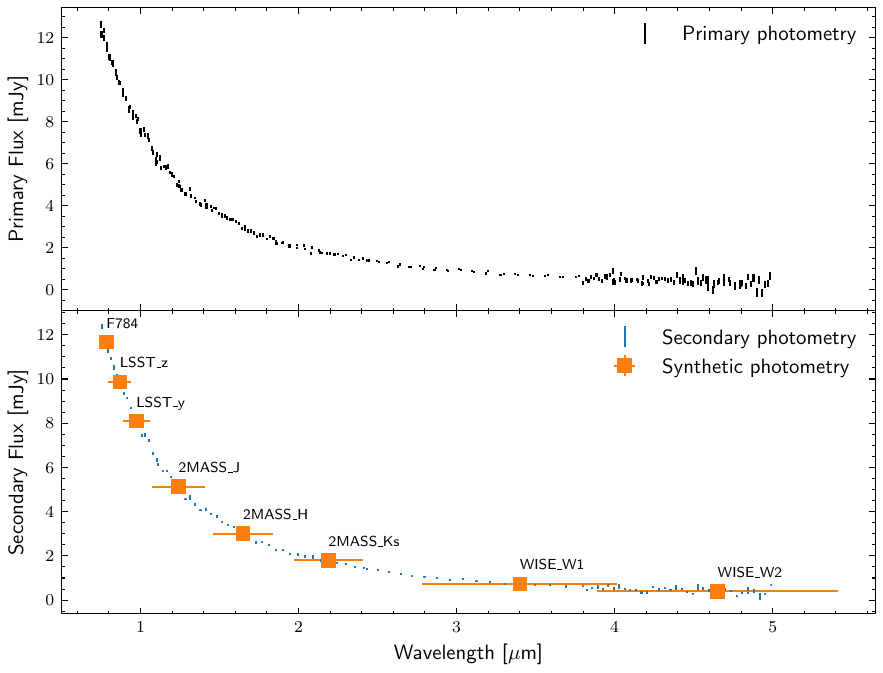}
\caption{An example of simulated source photometry using QuickCatalog for a primary calibration source (GD71).  Upper panel: simulated primary photometry (at the exact wavelength observed by \spherex).  Lower panel: the secondary catalog, containing photometry binned into 102 spectral channels (blue) and synthetic photometry (orange squares) where \spherex\ photometry data are binned into synthetic photometry for LSST, 2MASS and WISE bands. 
\label{fig:quickcatalog_fluxes}}
\end{figure}

The simulated photometric measurement is performed through ``optimal" forced photometry (meaning applying a weighting function to the pixel sum based on the PSF that maximizes the signal-to-noise of the photometry measurement) on a postage stamp image only containing a single isolated source.  Optionally, Tractor's PSF fitting photometry \citep{Lang2016} can be used for photometry in order to include the effects of confusion from neighboring sources.  In this mode, the postage stamp images are rendered with multiple nearby sources, a simple median background estimated, and Tractor is used to photometer all sources in the scene assuming exact priors on the source positions.
All observations of each source is then assembled into a catalog with the same format as the Primary Catalog.  This can further be reduced into the binned Secondary Catalog format, including synthetic photometry (see Fig.~\ref{fig:quickcatalog_fluxes}).

\section{Simulator Inputs}
\label{sec:inputs}
\begin{figure}
    \gridline{
        \fig{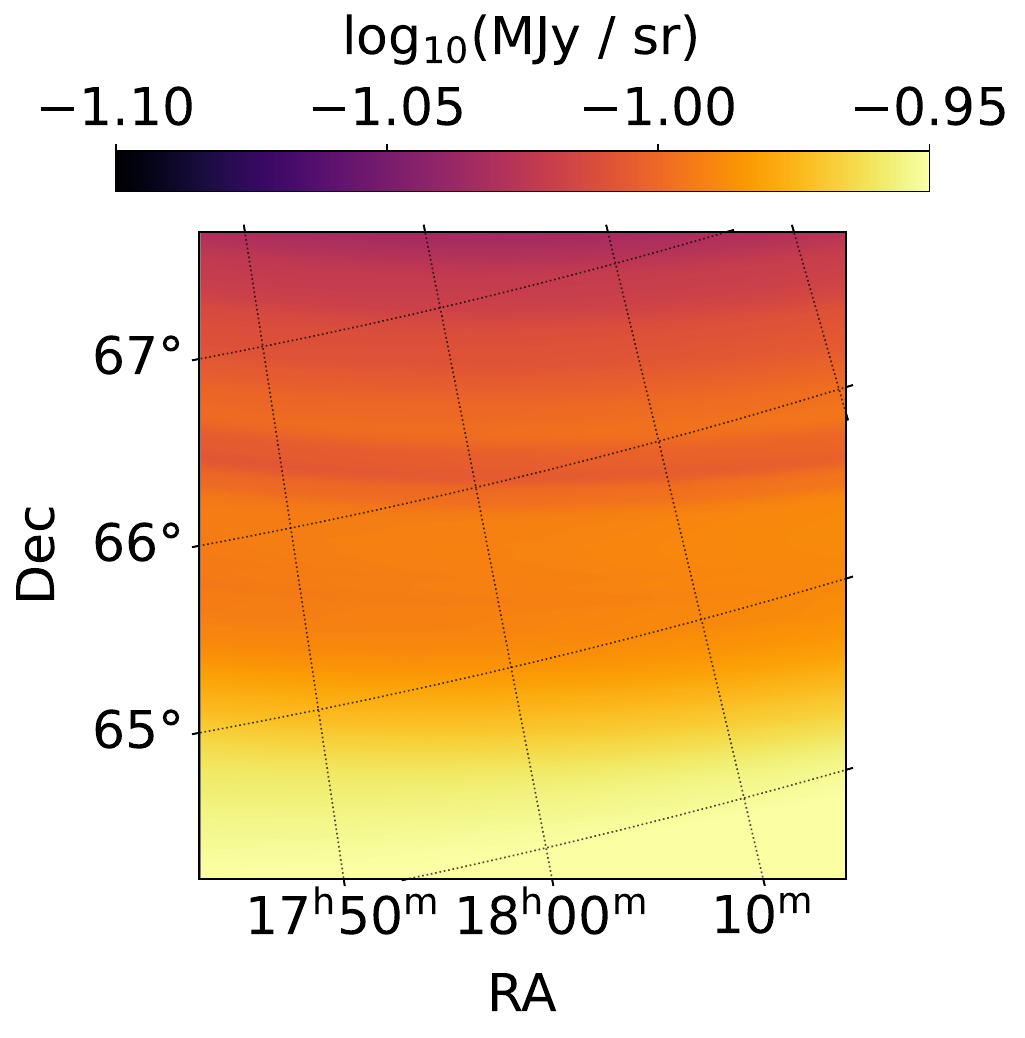}{0.2602065\textwidth}{Zodiacal light}
        \fig{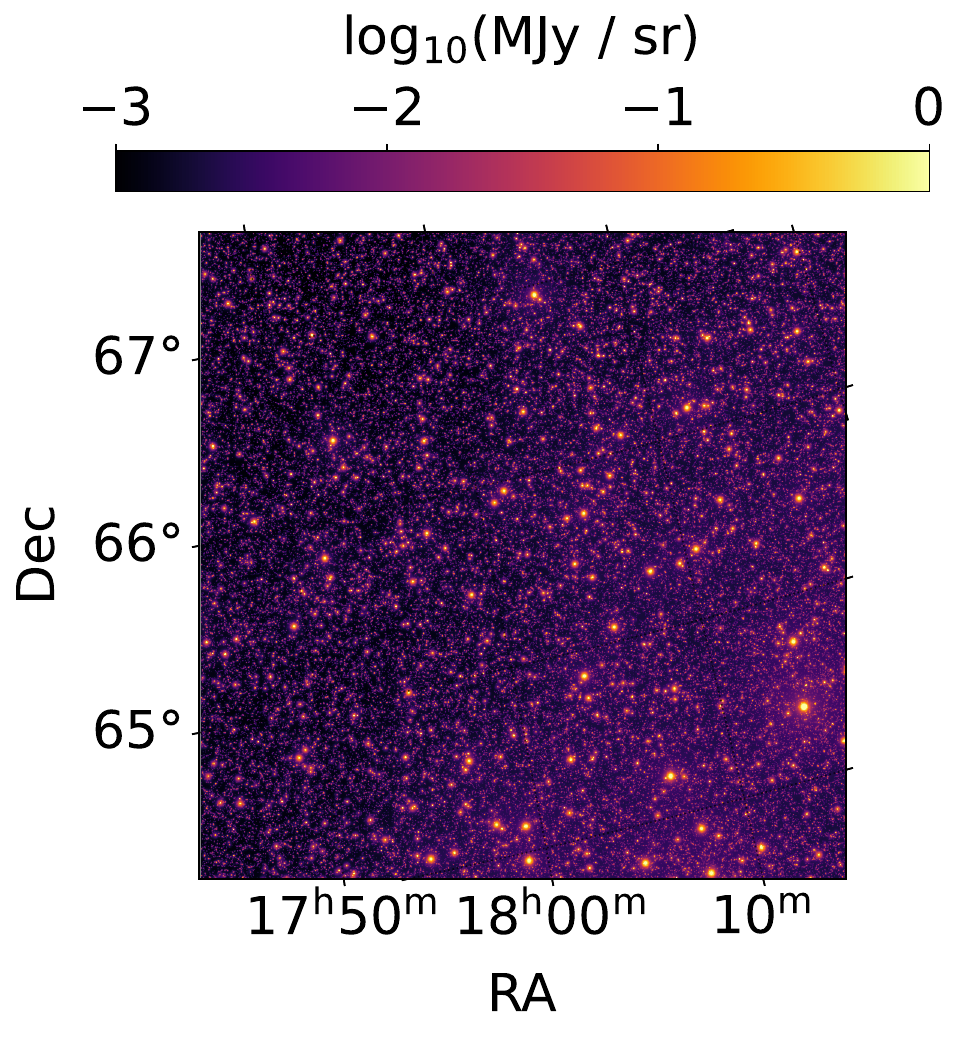}{0.2441938\textwidth}{Stars}
        \fig{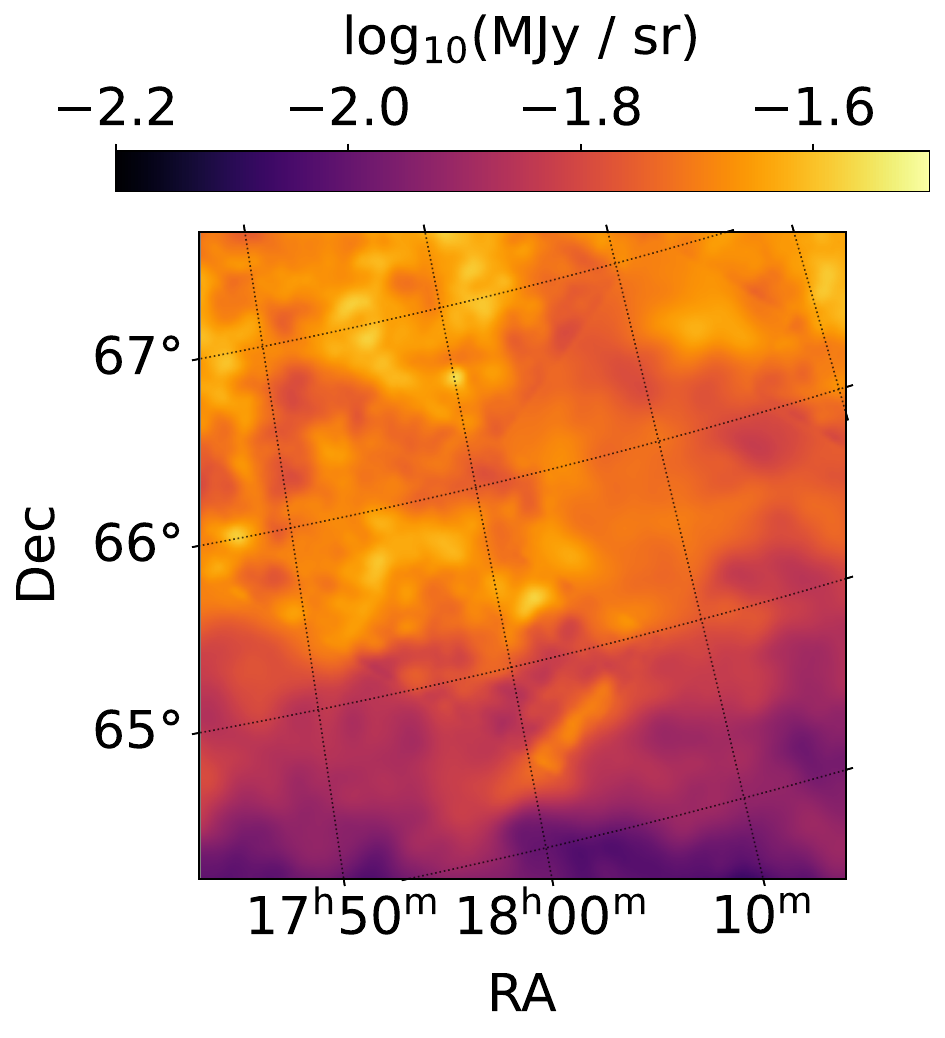}{0.24\textwidth}{Diffuse Galactic light}
        \fig{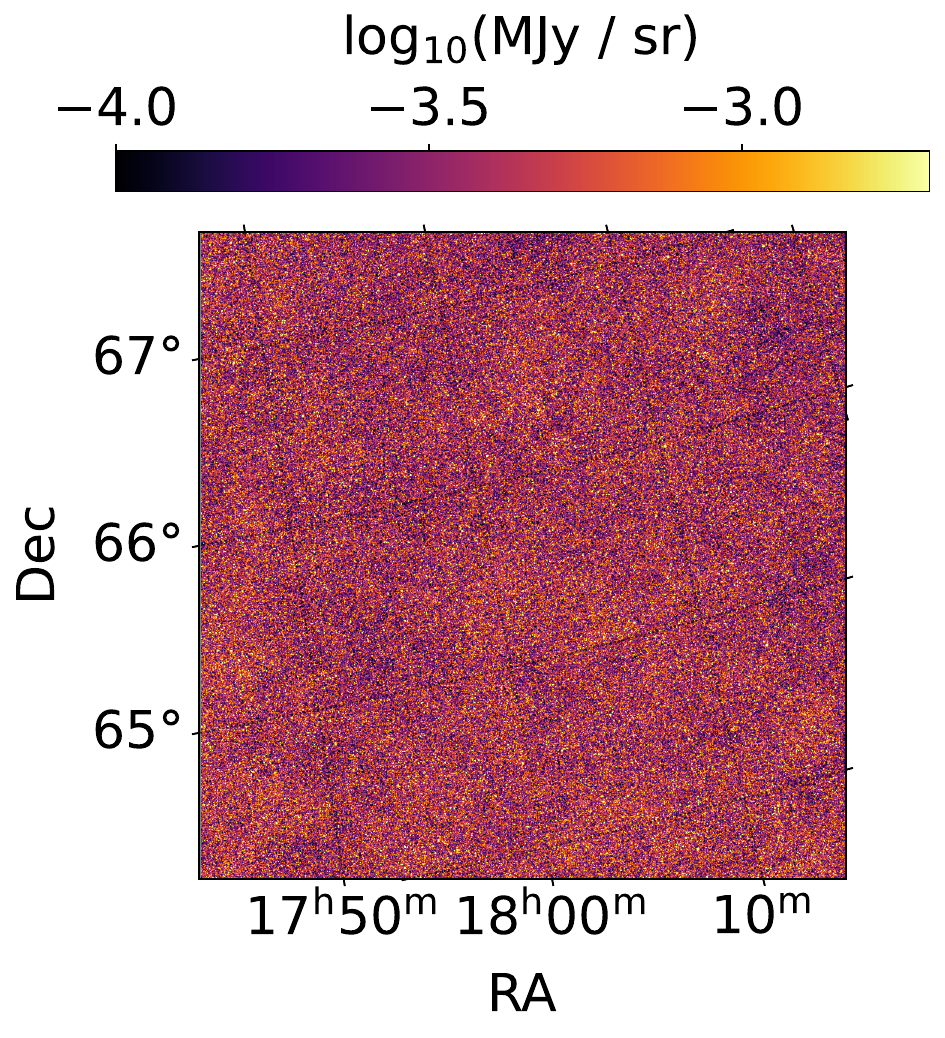}{0.24\textwidth}{Extragalactic background light}
    }
    \gridline{
        \fig{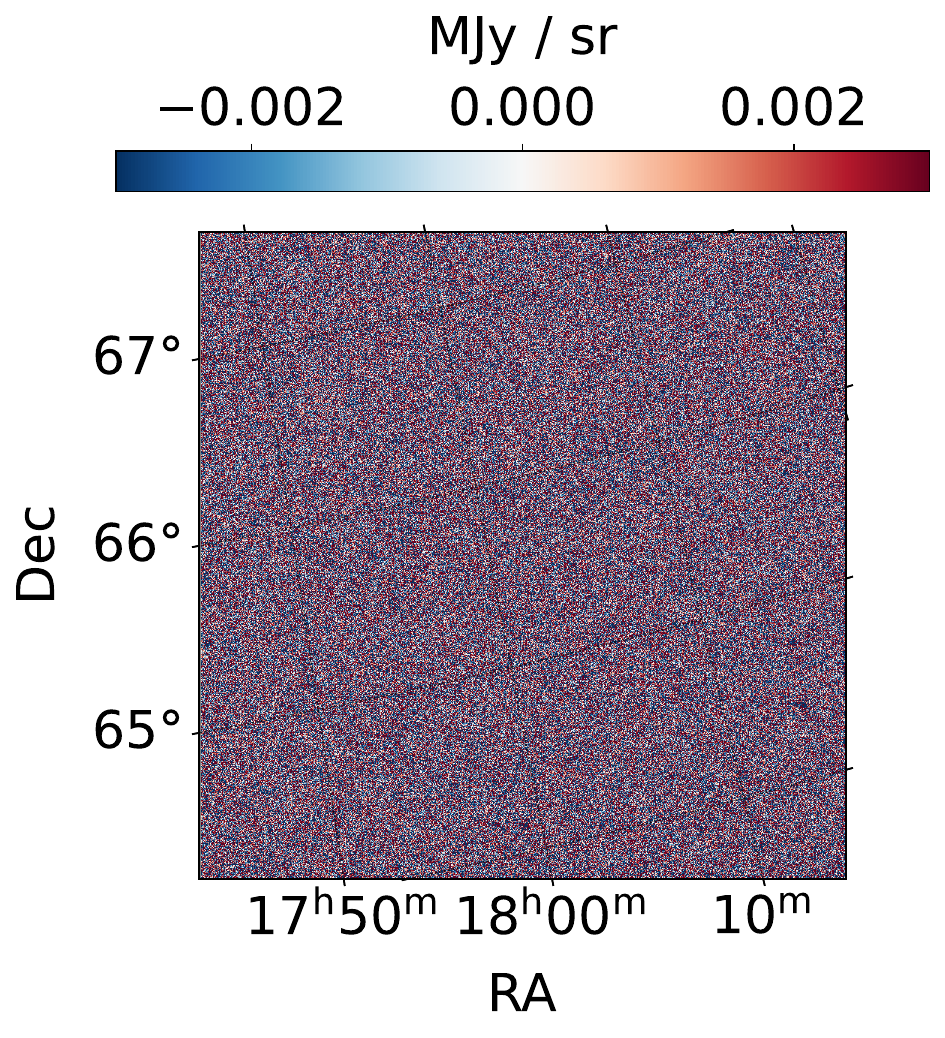}{0.24\textwidth}{Photon noise}        \fig{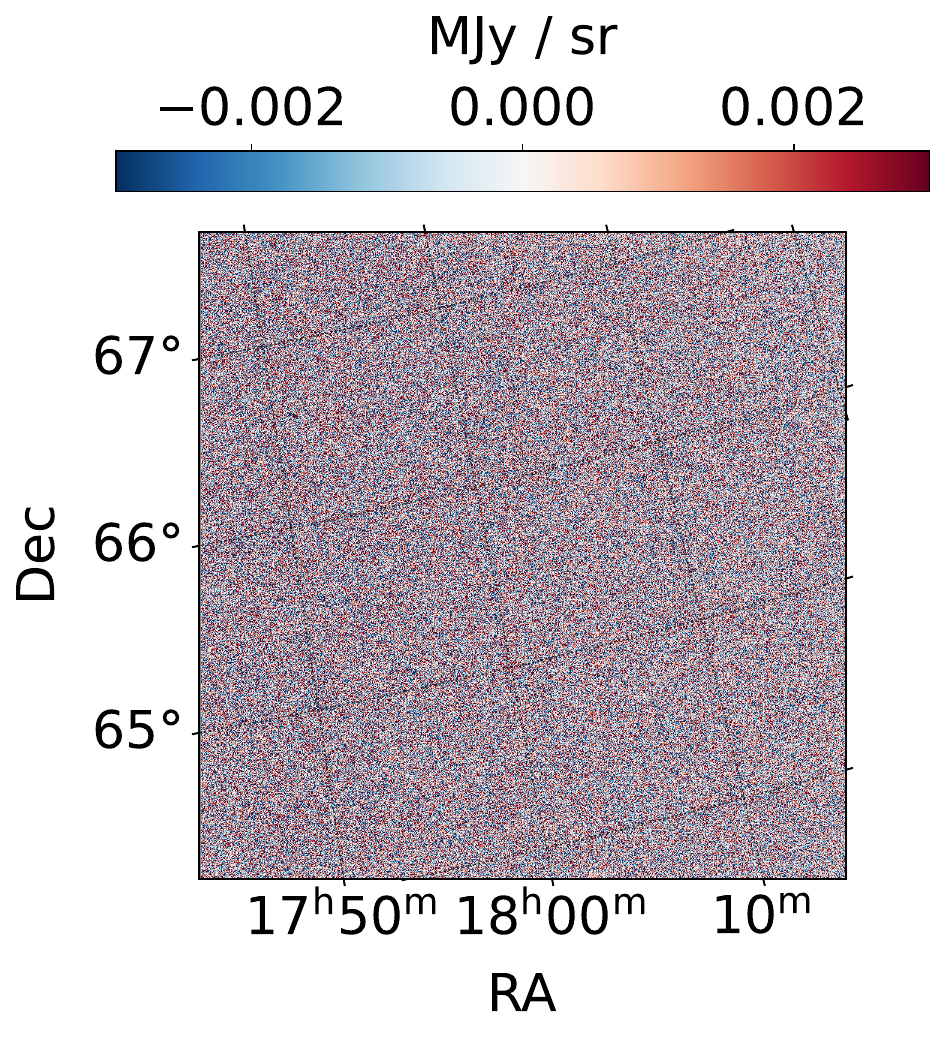}{0.24\textwidth}{Read noise}
        \fig{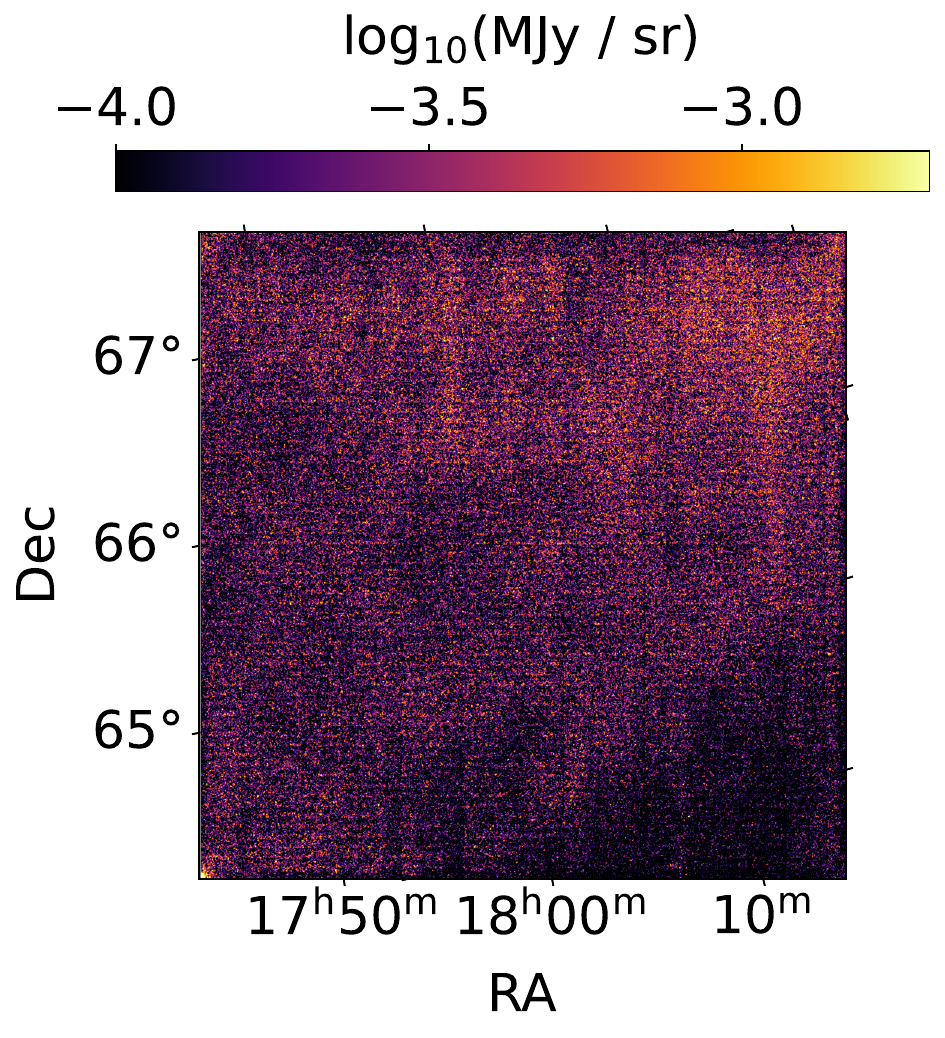}{0.24\textwidth}{Dark current}
    }
    \caption{Separated simulation components for array~1 for the exposure image shown in \figref{fig:exampleImages}. The component images can be simulated independently and then summed linearly to produce the full image. Individually resolved galaxies can be included in simulation of the extragalactic background light, or to simulate the true sky, galaxies can be instead drawn from the reference catalog. Photon noise is calculated from all optical components. Read noise and dark current are estimated from pre-launch laboratory testing.}
    \label{fig:simComponents}
\end{figure}

The primary purpose of the \skysim\ is to generate realistic images of the astronomical sky as it will be observed by \spherex\@.  Implementing high-fidelity models of the expected sources of astrophysical emission is important to understanding both the appearance of astrophysical sources in the images, as well as to accurately modeling downstream effects like photon noise.  In this section we describe the \skysim\ models for astrophysical sources including ZL, DGL, compact and extended galactic and extragalactic sources, artificial satellites around earth, and planets in our solar system. In \figref{fig:simComponents}, we present a breakdown of the major simulation components that contribute to each \spherex\ image.

\subsection{Zodiacal Model}

At optical and near-IR wavelengths, the brightest source of diffuse emission above the Earth's atmosphere is sunlight scattered and thermally emitted by interplanetary dust in the solar system, called ZL \citep{Leinert1998}. 
ZL is expected to be the dominant foreground in \spherex\ images, and must be accounted to calculate the photon noise and large scale structure of diffuse sky brightness in the simulations.  In the \skysim\ we implement a modified Kelsall model that builds on the original DIRBE-based model of \cite{Kelsall1998} as follows.  The standard Kelsall model applies geometry and scattering functions to different interplanetary dust populations to predict the specific intensity of the observed ZL as specific bands.  The \spherex-specific modifications incorporate the Solar spectrum\footnote{\href{https://www.nrel.gov/grid/solar-resource/spectra-astm-e490.html}{2000 ASTM Standard Extraterrestrial Spectrum Reference E-490-00}}, recent scattered-light measurements \citep{Tsumura2010,Tsumura2013_zodi,Matsumoto2015,Kawara2017}, and \textit{Planck} measurements of the emissivity of the interplanetary dust cloud from sub-mm observations \citep{planck2013-zodi}.
Additionally, the scattering phase function \citep{Hong1985} at the visible wavelength is utilized to compute the ZL at $0.55\,\mu\mathrm{m}$ and to interpolate the ZL at \spherex\ bandpass wavelengths shorter than $1.25\,\mu\mathrm{m}$, the shortest wavelength in the original Kelsall model.
The scattering phase function of \citet{Hong1985} exhibits anomalous behavior at scattering angles smaller than $\sim 30^\circ$ when the radial power-law exponent, $\alpha$, of the dust distribution exceeds $\sim 1$\@.
However, in the Kelsall model, $\alpha$ is set to 1.34\@.
To address this issue, we fitted a sum of three Henyey-Greenstein functions to the scattering phase function with $\alpha = 1.34$ at scattering angles greater than $30^\circ$ and adopted the fitted function in the \skysim.

The base model performs direct integration based on the \spherex\ spacecraft's position within the solar system and the line of sight, as described in \cite{Kelsall1998}.  The addition of the complete solar spectrum permits calculating the ZL over the entire \spherex\ passband, and is normalized to agree with the Kelsall model at $1.25 \, \mu$m.  Small modifications to the scattered (thermal) light spectrum to better match recent measurements are implemented as small adjustments to the albedo (emissivity) of the dust population.  A slightly higher performance version was also implemented based on the method of \cite{maris2019} which uses pre-computed full sky lookup tables of the modified Kelsall model.
\citet{maris2019} expressed the time dependence of ZL using a Fourier series as a function of time, whereas we adopted the Earth's mean longitude as the basis for the Fourier series expansion.
As a result, the lookup tables remain independent of the mission period and exhibit a periodicity of one year.
Additionally, to balance performance and accuracy, we set the maximum Fourier order to 6 instead of the original value of 8.

\begin{figure}[ht!]
\centering
\includegraphics[width=0.5\textwidth]{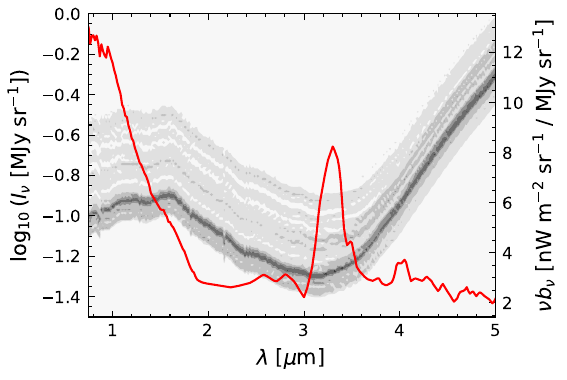}
\caption{Spectral energy distributions of the ZL and DGL used in the \skysim.  The grey color scale referenced to the left hand axis shows the surface brightness density computed over 256 random draws from the modified Kelsall ZL model covering the entire 2-year survey.  There is a factor $\sim 3$ brightness variation between the ecliptic plane and the poles, with smaller variations at longer wavelengths.  The red line for DGL referenced by the scale on the right hand axis shows the conversion factor $\nu b_{\nu}$ used to convert the far-infrared brightness to surface brightness in the near-infrared \citep{Sano2016}.  The conversion is brightest at optical wavelengths and shows a broad feature between 3 and 3.5~$\mu$m dues to polycyclic aromatic hydrocarbons in the interstellar medium.  At high galactic latitudes the far-infrared emission typically has brightness $\sim 1 \,$MJy sr$^{-1}$ in these units, which multiplying by $\nu b_{\nu}$ leads to a signal that is typically $\sim$10~\nw\ in the near-infrared.  This is over an order of magnitude fainter than the ZL.
\label{fig:foreground_sed}}
\end{figure}

The spectral energy distribution (SED) of the ZL model is illustrated in Fig.~\ref{fig:foreground_sed}, where we calculate the surface brightness for 256 observations drawn at random over the period of the two-year survey.  Various spectral features are evident, as is the factor of $\sim 3$ brightness variation in different regions over the sky.  

In the upper left panel of \figref{fig:simComponents}, we include an image of the typical zodiacal-light contribution to a \spherex\ exposure. Because zodiacal light is spatially smooth, the LVF image is dominated by the zodiacal SED, which produces features in the vertical~(wavelength) direction.

\subsection{Diffuse Galactic Light}

A model of DGL based on Planck and IRAS measurements of dust~\citep{planck_dust} scaled to the near-IR~\citep{Lillie1976,Draine2003,Zubko2004,Tsumura2013_DGL,Arai2015} is also computed by the \skysim.  The details of the spatial part of the model construction are given in Section 4.3.1 of \citet{Symons2023}.  The spectral component is modeled using the best-fitting $\nu b(\nu)$ model from the compilation of measurements in \citet{Sano2016}, reproduced in Fig.~\ref{fig:foreground_sed} here.  The DGL model has a fixed spectrum and does not include spatial features with characteristic sizes $\lesssim 5^{\prime}$, but gives a statistically plausible representation of the characteristics of the spatial and spectral intensity we expect to observe.   

In the upper row of \figref{fig:simComponents}, we include an image of the typical DGL contribution to a \spherex\ exposure. The DGL is dimmer than zodiacal light but has more significant spatial structure. At the wavelengths of the example in \figref{fig:simComponents}, the DGL is dominated by starlight scattered from interstellar dust grains, and this process is more efficient at shorter wavelengths. The LVF-induced wavelength variation creates a rough vertical gradient, whereby the upper portion of the DGL image is generally brighter than the lower portion.

\subsection{Compact Sources}
\label{sec:compactsources}

The \spherex\ Reference Catalog serves as an essential input for the \skysim, as well as for forced photometry of the \spherex\ data (Tractor) and other Level 1–3 software pipeline modules. This all-sky catalog is compiled from multiple external surveys, including Gaia \citep{GAIADR2,GAIAEDR3,Catalog_Gaia}, Pan-STARRS1 \citep{Flewelling2020,Catalog_PS1}, the DESI Legacy Imaging Survey \citep{Dey2019}, 2MASS \citep{Catalog_2MASS,Skrutskie2006}, AllWISE \citep{wright10,Mainzer2011,Catalog_Allwise}, and CatWISE \citep{Eisenhardt2020,Catalog_CatWISE, Marocco2021}\@.  The catalog contains approximately 5.8 billion sources, of which 1.1 billion are anticipated to exceed the $\sim10\sigma$ detection threshold. For further details, readers are referred to \cite{REFCAT_PAPER}. Below, we briefly describe the catalog components relevant to the \skysim.

The Reference Catalog provides SEDs and shape parameters for extended galaxies, which are crucial for simulating fluxes at the pixel level in the \spherex\ detectors. The \skysim\ utilizes three types of SEDs from the Reference Catalog: 1. flux tables, 2. single high-resolution SED templates, and 3. linear combinations of high-resolution templates.  By default, the \skysim\ estimates LVF fluxes using piecewise linear interpolation of broadband photometry provided in the Reference Catalog. The catalog includes photometric fluxes from Gaia ($G$, $G_{BP}$, $G_{RP}$), Pan-STARRS ({\sl grizy}), DESI Imaging Survey ({\sl grz}), 2MASS ({\sl JHKs}), and WISE ({W1/W2}). Additionally, various high-resolution spectral templates are available in the \spherex\ package, with the Reference Catalog specifying information on how to construct SEDs from these templates.  SEDs derived from templates can be parameterized by input redshift, template ID, normalization, $E(B-V)$, and ID for chosen reddening law. These high-resolution templates are used for simulating 9,953 ice sources \citep[SPLICES;][]{Ashby2023}, 15,703 galaxies from the COSMOS2020 catalog, and 98 CALSPEC calibrators \citep{Bohlin2014,Bohlin2017,Bohlin2022}. For more flexible galaxy SED modeling, \spherex\ can combine multiple template coefficients with input redshifts to generate composite SEDs. One example application of this method is integrating outputs from the EAZY photo-z code \citep{Brammer2008}, which employs 11 FSPS templates.  The \spherex\ bandpasses (Sect.~\ref{sec:bandpass}) are measured by photon counting detectors, so integrating the SEDs in spectral flux density units requires appropriate weighting, typically a $1/\lambda$ weighting.

In the upper row of \figref{fig:simComponents}, we include an image of the typical starlight contribution to a \spherex\ exposure. This simulation is produced through a combination of our Reference Catalog and our PSF~model~(\secref{sec:PSF}). Starlight varies over several order of magnitude. A small population of bright stars contributes signal far above the level of zodiacal light but only in localized regions. A much larger population of dimmer stars contributes at a moderate level across the entire image.

\subsection{High-resolution Extended Signals}
\label{sec:highresextsig}

Some sky signals are neither entirely diffuse nor entirely compact. In these cases, we must retain the finest scales and simulate PSF~convolution~(\secref{sec:PSF}), but we cannot draw the fluxes from a catalog as we described for compact sources~(\secref{sec:compactsources}). Instead, the input signals are provided as high-resolution maps at a chosen set of wavelengths. The wavelengths should be sufficiently dense to capture important spectral features. When handling the finest scales, it is impractical to construct input maps that cover the full sky. Instead, the maps cover restricted regions in a Cartesian projection. The \skysim\ tool that handles these inputs is referred to as the \emph{flat-sky calculator}.

\begin{figure}
    \gridline{
        \fig{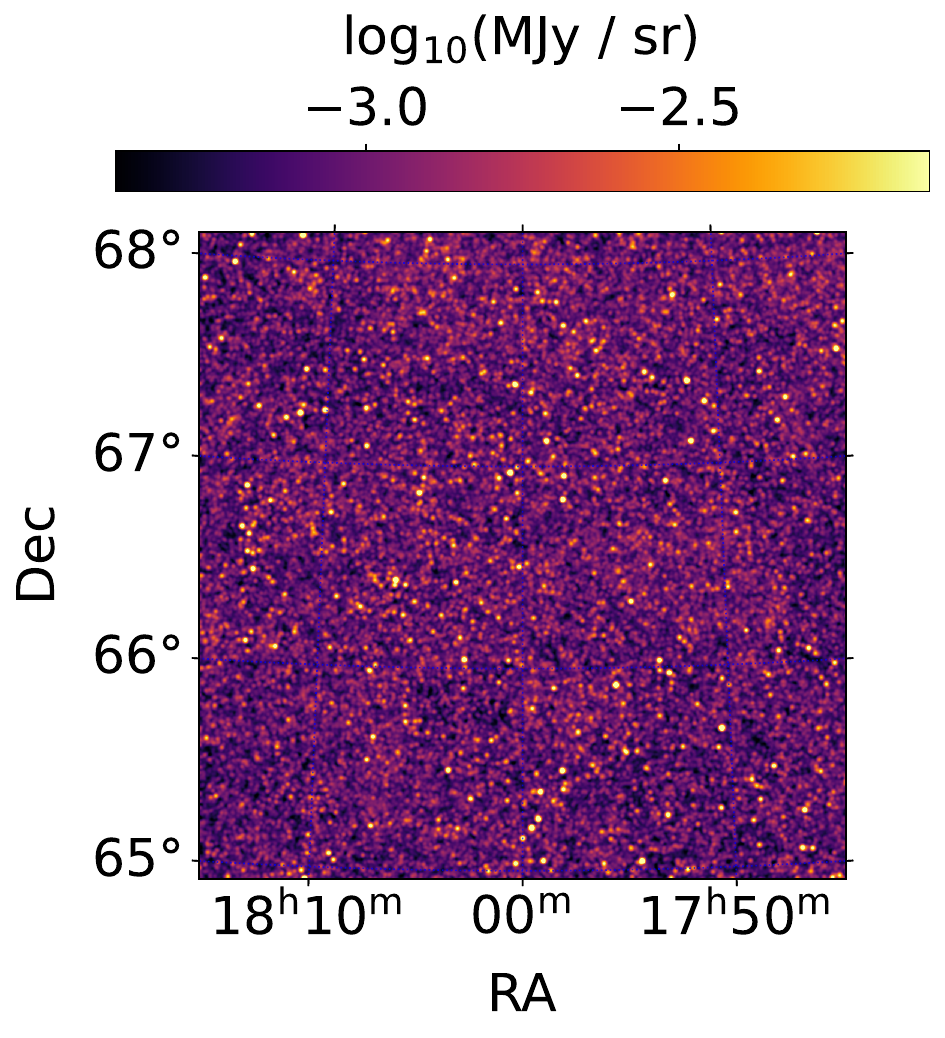}{0.24\textwidth}{$0.74~\mu\mathrm{m}$}
        \fig{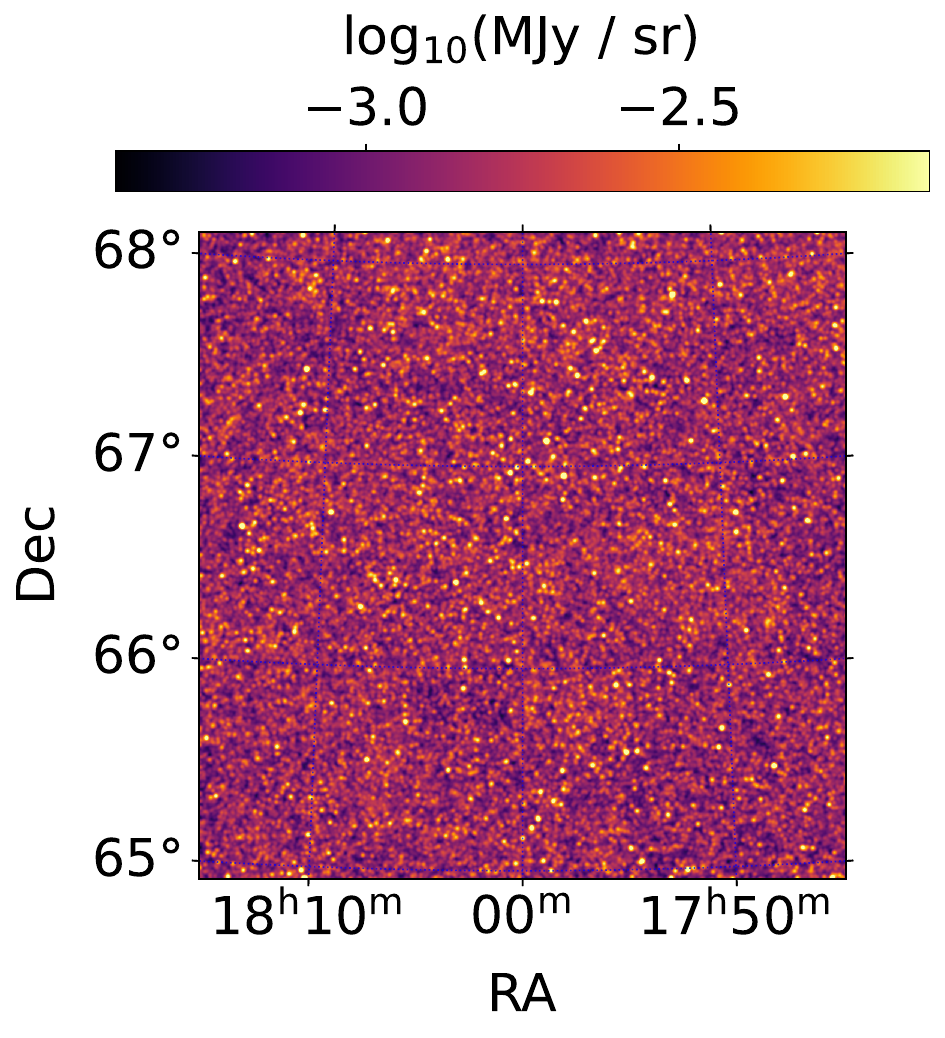}{0.24\textwidth}{$1.62~\mu\mathrm{m}$}
        \fig{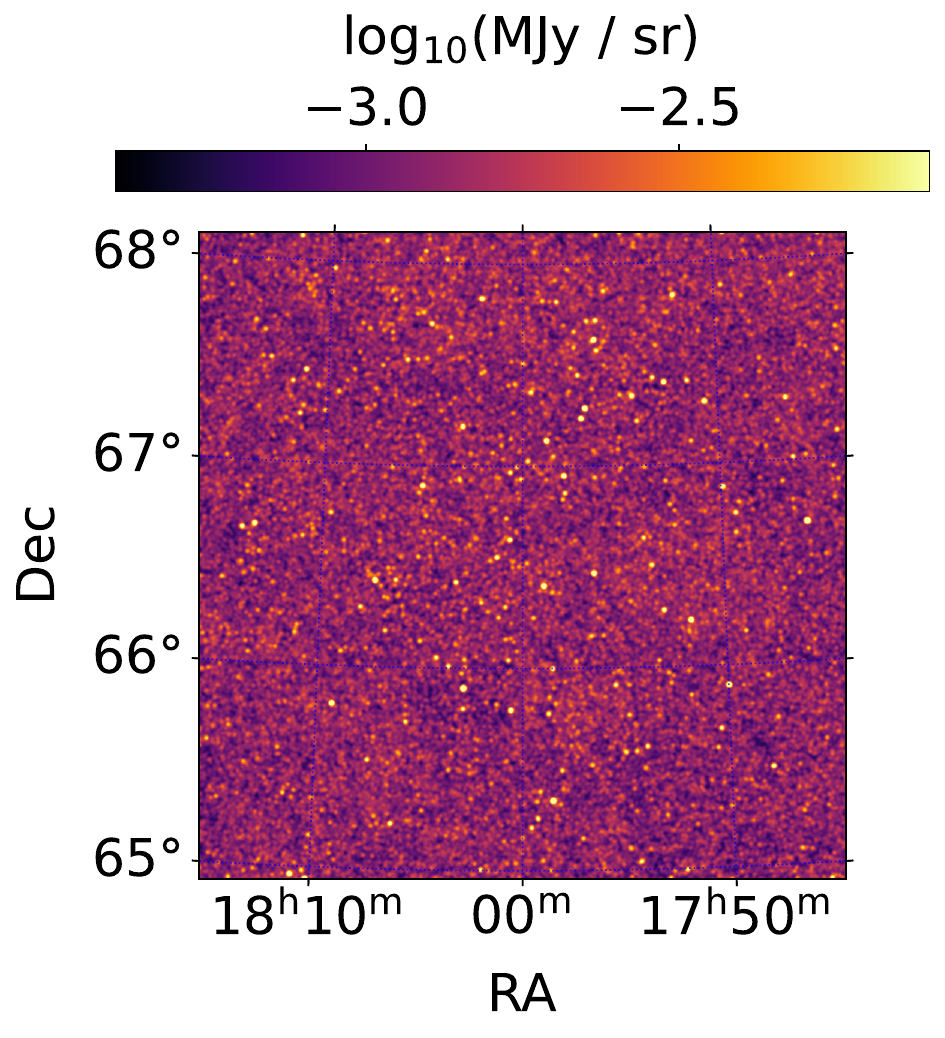}{0.24\textwidth}{$3.81~\mu\mathrm{m}$}
        \fig{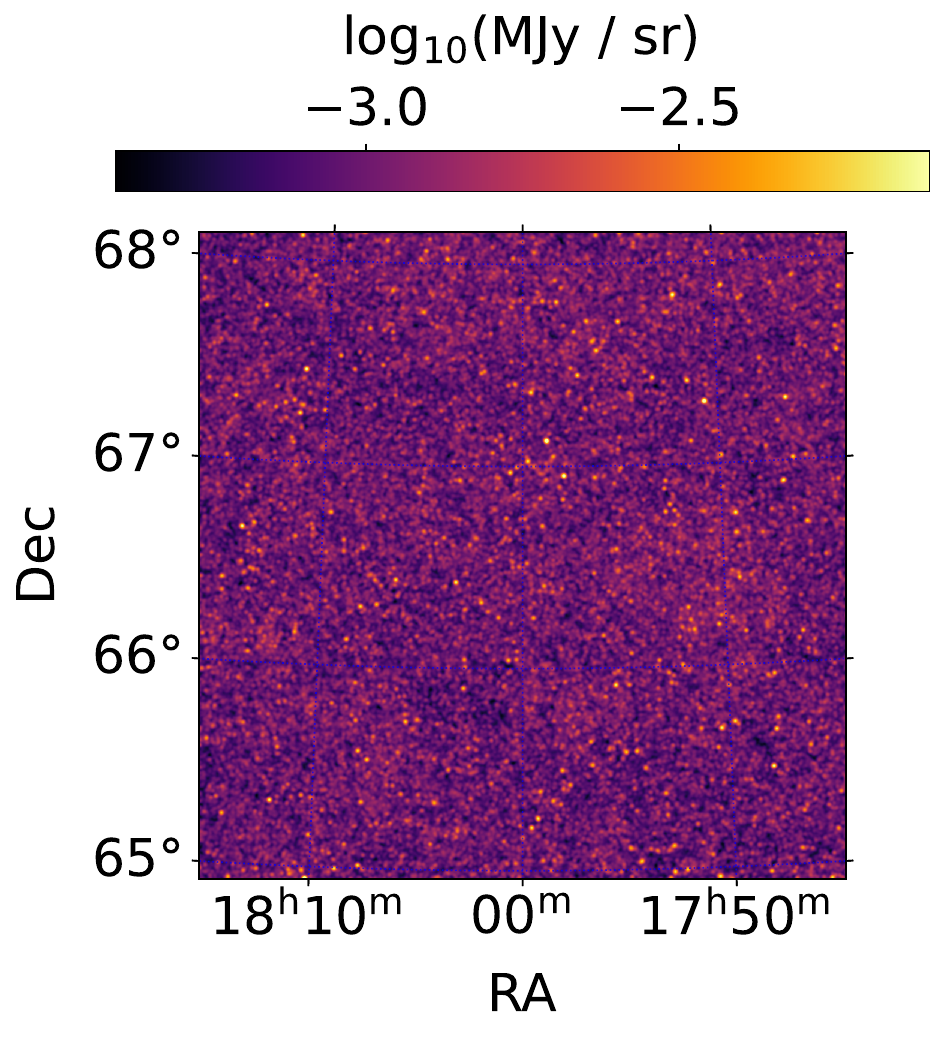}{0.24\textwidth}{$5.01~\mu\mathrm{m}$}
    }
    \caption{Simulated maps of extragalactic background light~(EBL). These are used as inputs to the \skysim. The native resolution of these maps is~$1\arcsec$. To better visualize the large-scale structure, we have here smoothed the maps to~$1\arcmin$. A full simulation set typically contains more than~100 maps in order to sample the \spherex\ wavelength range at the spectral resolution of the instrument. We have here selected a smaller sample of representative wavelengths. The above maps cover an area of~$3\deg \times 3\deg$ near the north ecliptic pole, which is the center of one of \spherex's deep fields. The simulated maps can be constructed over larger areas and translated to any region of sky. These simulations are accompanied by galaxy catalogs with simulated photometry to emulate the \spherex\ reference catalog.}
    \label{fig:EBLmocks}
\end{figure}

An important example is that of the extragalactic background light (EBL), which is the main observable for the \spherex\ investigation of the history of galaxy formation. The EBL consists of all of the light produced over cosmic time from outside of the Milky Way. This includes bright, nearby galaxies that can be detected individually; some may be resolved by \spherex, though most will appear as point sources. The EBL also contains the diffuse emission accumulated from the large population of sources that are too faint to be detected individually; these include dwarf galaxies, intrahalo light~(IHL) and distant, high-redshift galaxies. For the \spherex\ galaxy-formation study, EBL is simulated through the use of the halo model of~\cite{Mirocha2025}. 
Maps are produced for each of the 102~nominal \spherex\ wavelengths, though they can also be made more finely if needed. The spatial resolution is typically~$1$-$3\arcsec$, and the total area is typically no larger than $20^\circ \times 20^\circ$. The resolution is chosen to be fine enough to be able to simulate PSF~convolution, and the area is large enough to cover the \spherex\ deep fields, which are the main targets of the EBL investigation. 

In \figref{fig:EBLmocks}, we present a selection of representative EBL input maps. These belong to a larger data cube that informs the EBL contribution to simulated images. An example of an EBL image is shown in the upper right of \figref{fig:simComponents}. The LVF imparts a subtle vertical gradient that, for the wavelength range of the band in \figref{fig:simComponents}, enhances the EBL brightness toward the bottom of the image. Because each \spherex\ band covers 17~spectral channels, each EBL image is typically informed by at least 17 EBL input maps, where a larger number may be required to handle the edges of the band or to simulate finer spectral features.

Because the EBL also contains bright galaxies, the mock sky realizations are accompanied by mock catalogs that can be used to simulate source masking or to simulate cross-correlation studies. The catalogs can be modified to mimic the statistics of existing datasets such as that of DESI \citep{DESIDR1} or of future datasets such as that of \spherex\ itself.

\begin{figure}
    \gridline{
    \fig{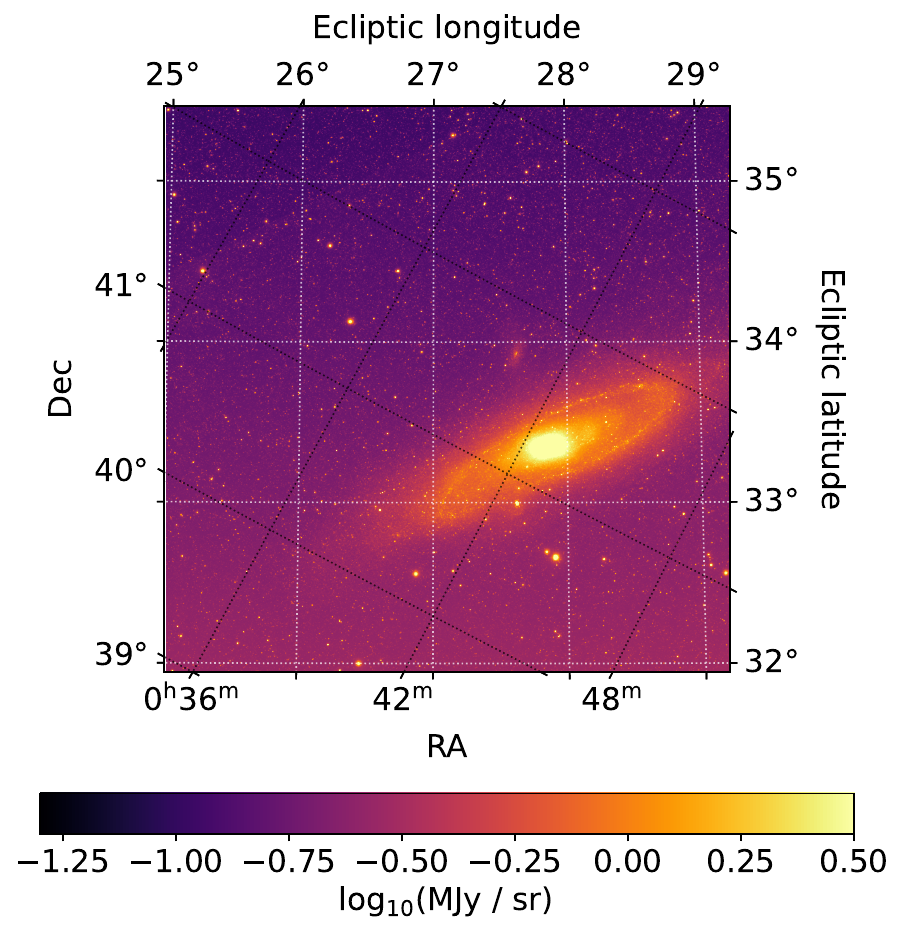}{0.4\textwidth}{Andromeda Galaxy~(M31)}
    \fig{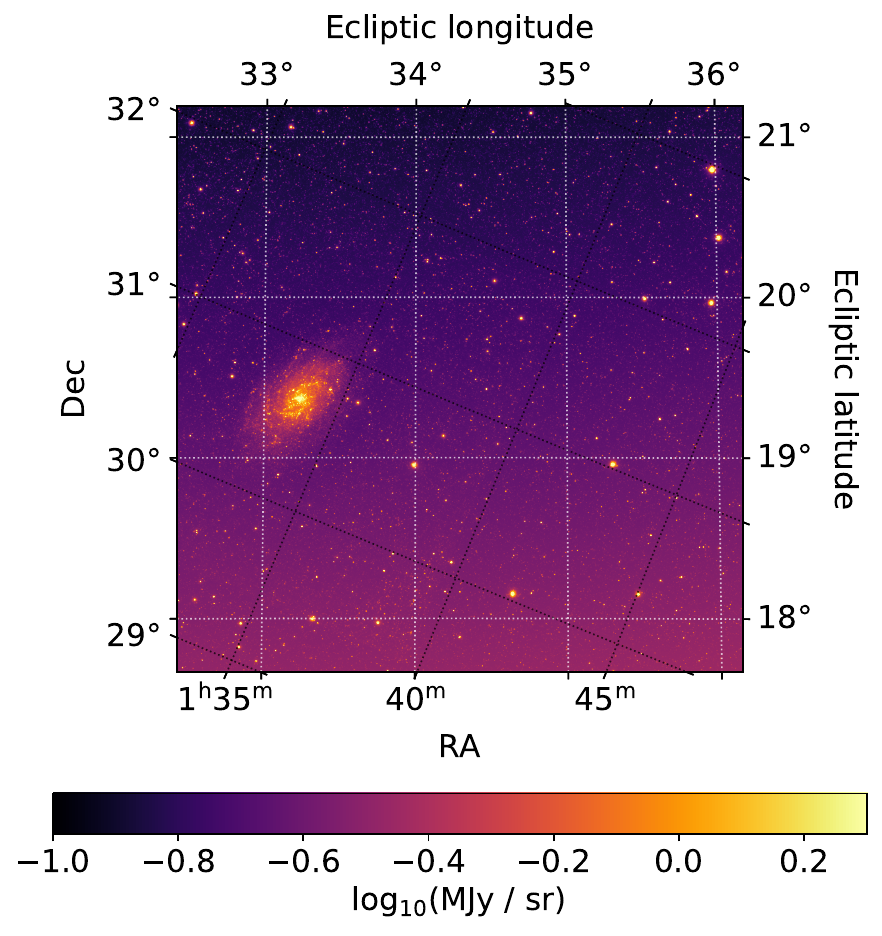}{0.39004\textwidth}{Triangulum Galaxy~(M33)}
    }
    \caption{Simulated images of extended sources observed with band~5. These images include all of the components from \figref{fig:simComponents} with the addition of extended sources drawn from the WISE Extended Source Catalog. Because these example objects are relatively bright, the background is dominated by stars and zodiacal light, where the latter produces a prominent vertical gradient in LVF images.  The black dotted grid corresponds to ecliptic coordinates while the white dotted grid corresponds to ICRS coordinates.}
    \label{fig:extendedSources}
\end{figure}

Another example is the population of extended sources, e.g., well-resolved galaxies and nebulae. These objects typically have complex morphologies with features on scales as fine as the \spherex\ PSFs and as large as the \spherex\ field of view. In principle, inputs can be constructed for objects that are real or synthetic. Such simulations may also serve as useful guidance when deciding how to mask extended sources. 

We have incorporated maps of the Andromeda Galaxy~(M31) and the Triangulum Galaxy~(M33) from the WISE Extended Source Catalog~\citep{Jarrett2019} as demonstrations, where bright point sources have been removed. These maps are provided only for the relatively broad bands of WISE, so \spherex-scale spectral features are largely obscured. Nevertheless, the test demonstrates a key capability of the \skysim\ and allows for estimates of expected sensitivity for similar objects. Simulated images of~M31 and~M33 are presented in \figref{fig:extendedSources} along with all of the simulation components from \figref{fig:simComponents}. Because these galaxies are relatively bright, the background features are dominated by starlight and zodiacal light. The LVF imparts a vertical gradient on all components including~M31 and~M33.  

\subsection{Earth Satellites}

Low-Earth Orbit satellite megaconstellations are known to have increasing impacts on terrestrial astronomy observations in the visible and near-infrared \citep[e.g.,][]{Mroz2022}\@.  Despite \spherex's vantage point from a 650~km low-Earth orbit,
Earth satellites at higher orbits will cross the field-of-view of \spherex.  Due to the rapid relative velocity with respect to \spherex, these will leave streaks in the exposures.  The streaks will be detected and flagged in the analysis pipeline.

To estimate the brightness and spectrum of Earth satellites, the \skysim\ uses a model that follows \cite{Hainaut_2020} in the visible as a diffuse scatterer of sunlight; and a blackbody emitter in the near-infrared \citep{JASON_2020}.  The relative velocity and range between \spherex\ and the source are drawn from values from these references.    Based on the known distribution of satellite orbits, typically 1-2 satellite streaks will be observed per image, with slightly more when the spacecraft is above the Earth's equator. 

\begin{figure}[ht!]
\centering
\includegraphics[width=\textwidth]{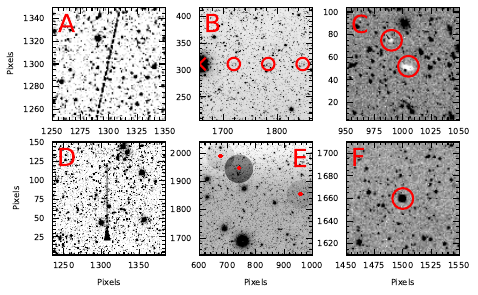}
\caption{``Rogues' Gallery'' of various local and instrumental features modeled by the \skysim.  These images are drawn from a variety of bands and sky positions and are meant to show typical examples of each effect.  The photocurrent scaling varies between images, but typically covers a range of about 1 \eps.  (A) An example of a signal expected from a Earth satellite crossing the \spherex\ field of view.  This $100\times 100$ pixel cut out captures a small part of a streak that crosses the entire detector.  (B) Electrical crosstalk, where the primary source - a bright star at left - is marked with an `x' and the first three electrical crosstalk images are circled.  We have exaggerated the effect by $25 \times$ in amplitude to highlight its structure.  (C) Flat-field imperfections due to detector artifacts, with some examples circled.  The simulated images include responsivity variations that, in addition to large scale structures, can cause localized regions of poor photoresponse that are modeled by the \skysim.  (D) Frame edge ghost, which is due to light from bright stars just off the image to scatter off the frame that holds the filter above the detector array.  This causes roughly linear features perpendicular to the edge of the image.  (E) Diffuse beam splitter ghosts caused by internal reflections in the optical system.  These are large circular features (centered on the `+' marks) that are sourced by bright stars in certain geometries compared with the focal plane. (F) Detector image persistence, an electrical effect where the charge caused by the images of bright sources are not fully flushed from the detector material after a well reset.  The circled source is spurious and was caused by a bright star in a previous exposure whose image has not been fully erased by the reset.  These images look similar to stars and decay away with a time constant of several hundred seconds. 
\label{fig:roguesgallery}}
\end{figure}

To highlight this and other effects that are visible in the simulated photocurrent images, in Fig.~\ref{fig:roguesgallery} we provide a ``rogues' gallery'' to illustrate various features.  Typical satellite streaks fall across the entire detector array, causing linear features that are clearly distinguishable from astronomical emission.

\subsection{Bright Planets}

Five bright planets (Mars, Jupiter, Saturn, Uranus, and Neptune) are available for detailed modeling in the \skysim. Their near-infrared spectra are modeled using the Planetary and Universal Model of Atmospheric Scattering (PUMAS, \citealt{2018JQSRT.217...86V, 2022fpsg.book.....V}) with the single-scattering radiative transfer method. The true fluxes at a given timestamp are scaled based on the positions of the planets relative to the geocenter, using the ephemeris generated by the planetary ephemeris DE432s \citep{Folkner_2014}, as it is sufficiently accurate and small in size (10 MB). The phase correction is applied based on the model presented in \citet{2018A&C....25...10M}, assuming no wavelength dependency. Since planets can have large angular sizes, they are treated as two-dimensional ellipses using real equatorial and polar radii \citep{2007CeMDA..98..155S}. Specifically, for Saturn, three rings (A, B, and C) are modeled, and their fluxes are added to that of Saturn (the C-ring's outer radius is set as the inner radius of the B-ring).

Three of the bright planets are  among the brightest objects that \spherex\ will observe;   typical apparent V-band magnitudes are $0.5$ AB for Mars, $-2.2$ AB for Jupiter, and $1.0$ AB for Saturn.  Each of these sources will appear in 300 to 500 images out of approximately 2.3 million exposures taken during the 2-year mission.

Based on the instrument model described below, Mars, Jupiter, Saturn, and Uranus will saturate multiple detector pixels given the fixed exposure time of the survey.
In addition, scattered light from the planets will be seen in the wings of the extended PSF, even when the source is several degrees off-axis.  
The exact behavior of the extended PSF will be studied once flight data becomes available, both through stacking signals from bright sources and response to the Moon outside the field of view during in-orbit checkout.
We include these bright planets optionally in simulated images to test the performance.


\section{Survey Planning}
\label{sec:survey}

Operating from low-earth orbit, \spherex\ observes in all 102 spectral channels, across the entire sky, and in two deep fields, all while staying within a set of time-varying geometric and operational constraints that define our allowable pointing zone. While many other observatories have faced their own optimal observing tasking problems, no existing observation planning software was suitable for our unique set of requirements. We therefore developed our own Survey Planning Software (SPS). We leveraged heritage from WISE \citep{wright10}, described our initial concept in \cite{Spangelo2015}, and a paper with a full description of the SPS is in preparation. Here we review key features as they relate to the \skysim.

The SPS plans observations that deliver the required all-sky and deep field coverage for \spherex. Among other capabilities, the SPS plans science observations, pauses observation for downlink, mitigates outages due to passage through the elevated radiation environment of the South Atlantic Anomaly \citep{Kurnosova1962}, plans Moon avoidance slews, and selects a safe place to point in rare cases when no science targets are observable. \spherex\ has a single main target list that includes targets from the all-sky survey as well as targets in both deep fields. All targets are organized into \textit{target groups}. To move to a desired target group, \spherex\ executes a \textit{large slew} maneuver. After the observatory settles at this attitude, \spherex\ takes a single exposure, then executes a \textit{small slew} maneuver to the next target in the group, and continues taking exposures followed by small slews until the target group is about to move out of our allowable pointing zone. The SPS then plans a large slew to the next target group, and observing continues in this way throughout our science operations. To plan these slews efficiently, the SPS selects the sequence of target groups to minimize the number of time-expensive large slews, and maximize the number of observations made within a target group before it moves out of the allowable pointing zone.

In operation, the \spherex\ team obtains an up-to-date orbit prediction based on recent onboard Global Positioning System (GPS) data. We then run the SPS with this orbit to plan the next several days of science observations, and generate a file that contains the planned attitude as a function of time of the spacecraft. We then uplink the commands and \spherex\ operates accordingly. We repeat this operations cadence throughout the mission.

Before launch, to simulate science data and other aspects of our mission, we conduct mission-long simulations of observations using the SPS with a nominal theoretical orbit. This generates a mission-long (i.e. 25-month) survey plan that yields the planned attitude as a function of time for our entire mission if it were to exactly follow this nominal theoretical orbit. This file is used by the \skysim\ to generate simulated science data of the observations that correspond to this survey plan. Among other effects, the observation times from the survey plan inform the ZL model and the image persistence model, and the orbital velocity informs the model of coordinate aberration (that the SPS nominally corrects for in its planning).  For each pointing, the spacecraft attitude and velocity in the survey plan also define the specific World Coordinate System (WCS) that transforms sky coordinates to detector pixel coordinates to enable mapping the simulated sky signal onto the \spherex\ detectors in a realistic way.

\section{Instrument Model}
\label{sec:instrument}
The \skysim\ includes a detailed model of the instrument to realistically reproduce how data are processed by its optical and electronic systems. The instrument model acts on the astrophysical signals returned by the sky model.  First, a model for the action of the optical system simulates the telescope transfer function including field distortion, the PSF, spectral filtering, beam splitter ghosts, and frame edge ghosts.  The optical to electrical transduction is simulated to convert between astrophysical signal and electrical photocurrent in the detector.  At the detector level, a variety of effects including readout and photon noise, charge diffusion, dark current, image persistence, electrical crosstalk and inter-pixel capacitance (IPC) are simulated.  The \skysim\ also provides realistic flagging of pixels as they are expected to be reported by the on-board software, but we do not simulate potential losses of data in the telemetry.

The behavior of the instrument at each of these steps was initially set based on the design of the mission and hardware components.  As these components were delivered and tested, data from component integration and test campaigns informed the \skysim.  Final thermal vacuum tests of the integrated telescope and detector system provided all the pre-launch parameters and characterization needed for realistic simulations.  
Per-detector parameters used in the \skysim\ instrument model are summarized in Table~\ref{tab:spherex_bands}.

\begin{table}[ht]

\begin{tabular}{ccccccccc} 
\hline
Band & $\lambda_{\mathrm{low}}$& $\lambda_{\mathrm{high}}$  & R & $\eta_{\mathrm{opt}}$ & $g$ & pixel scale & PSF FWHM & $\sigma_\mathrm{CDS}$\\
& [\micron] & [\micron] & & & (e$^{-}$ s$^{-1}$) / ($\mathrm{MJy} \cdot \mathrm{sr}^{-1}$) & [${}^{\arcsec}$] & [${}^{\arcsec}$] & [e$^{-}$]\\
\hline
1 & 0.73 & 1.13 & 39.1 &  0.81 & 8.29 & 6.14 & 4.0 & 17.2\\

2 & 1.08 & 1.67 & 41.4 &  0.83 & 8.39 & 6.12& 3.8 & 16.5\\

3 & 1.61 & 2.45 & 40.6 &  0.77 & 7.90 & 6.14& 3.6 & 17.2\\
 
4 & 2.38 & 3.87 & 34.8 &  0.81 & 9.87 & 6.15& 4.3 & 10.8\\

5 & 3.79 & 4.44 & 111.5 &  0.57 & 2.05 & 6.13 & 5.2 &  13.9\\

6 & 4.39 & 5.02 & 128.2 &  0.53 & 1.68 & 6.15& 5.5 & 11.6\\
\hline 
\end{tabular}
\caption{Per-detector parameters used in the \skysim\ instrument model.  The upper and lower wavelengths ($\lambda_{\mathrm{low}}$, $\lambda_{\mathrm{high}}$) of each detector show the range of all pixels in the detector arrays for each band, not only those that contribute to full spectral channels. Note that the bands by design includes padding of a quarter of a spectral resolution element.  $R$ is the median spectral resolution $\lambda$/$\Delta\lambda$ in each band. $\eta_{\mathrm{opt}}$ is the end-to-end detector-mean optical efficiency. $g$ is the conversion factor from astrophysical signal expressed in surface brightness units to photocurrent at the detector. The pixel scale is the mean over the detector array.  The quoted full-width-at-half max (FWHM) is the mean for the band and does not include pointing jitter.  $\sigma_\mathrm{CDS}$ is the 1$\sigma$ electronic readout noise per pixel, expressed for Correlated Double Sampling (difference of two reads).}
\label{tab:spherex_bands}

\end{table}

\subsection{Optical Modeling}

Light from the astrophysical sky is first processed by the optical chain whose action can be simulated at the component level.

\subsubsection{Optical Distortion}
The \spherex\ telescope is a three-mirror anastigmat with free-form mirrors optimized in shape to produce low wavefront error across a 3.5$^{\circ} \times$ 11.3$^{\circ}$ field of view.  Nonetheless, aberrations and field distortion across the field of view are significant. 

The in-flight data will use thousands of stars with Gaia astrometry seen in each exposure to solve for the field distortion as part of the fine astrometry calibration.   Before the flight data are available,  the \skysim\ uses a field distortion model; the pre-launch instrument test campaign did not include characterization of field distortion.    A ray-trace model of the as-built telescope was used to calculate a set of 81 ordered pairs of sky and focal plane coordinates for each of the six detectors.  The ordered pairs were fit to a model that includes a polynomial transformation from undistorted to distorted pixel coordinates, rotation and scaling of the focal plane, and a tangent plane projection onto the sky \citep{gwcs}.  A third-order polynomial transformation provided a fit with an rms residual of better than 10$^{-3}$ of a pixel.   Higher order polynomials did not improve residuals further.  The model predicts distortions up to 40 \spherex\ pixels as compared to an ideal undistorted telescope in the far corners of the field of view.  The model also determined the mean pixel plate scale and variation of the projected pixel solid angle across the field of view. 

This model is used in the \skysim\ to create the WCS relationship between sky coordinates and pixel coordinates, needed to place the astrophysical source signals onto detector pixels.
The optical distortion is represented by Simple Imaging Polynomial~(SIP) coefficients~\citep{FITS_SIP} in the FITS header.

\subsubsection{Point Spread Function}
\label{sec:PSF}

The as-built model of the telescope used for the optical distortion model is also used in optical modeling software to generate a library of PSFs across the field of view, taking into account the variation of wavelength response across each detector.  This work was performed by Photon Engineering\footnote{\url{https://www.photonengr.com/}} using their FRED ray-tracing software, which includes diffraction, internal reflection in the dielectric elements, and scattering effects.

\begin{figure}
    \centering
    \includegraphics[width=0.45\textwidth]{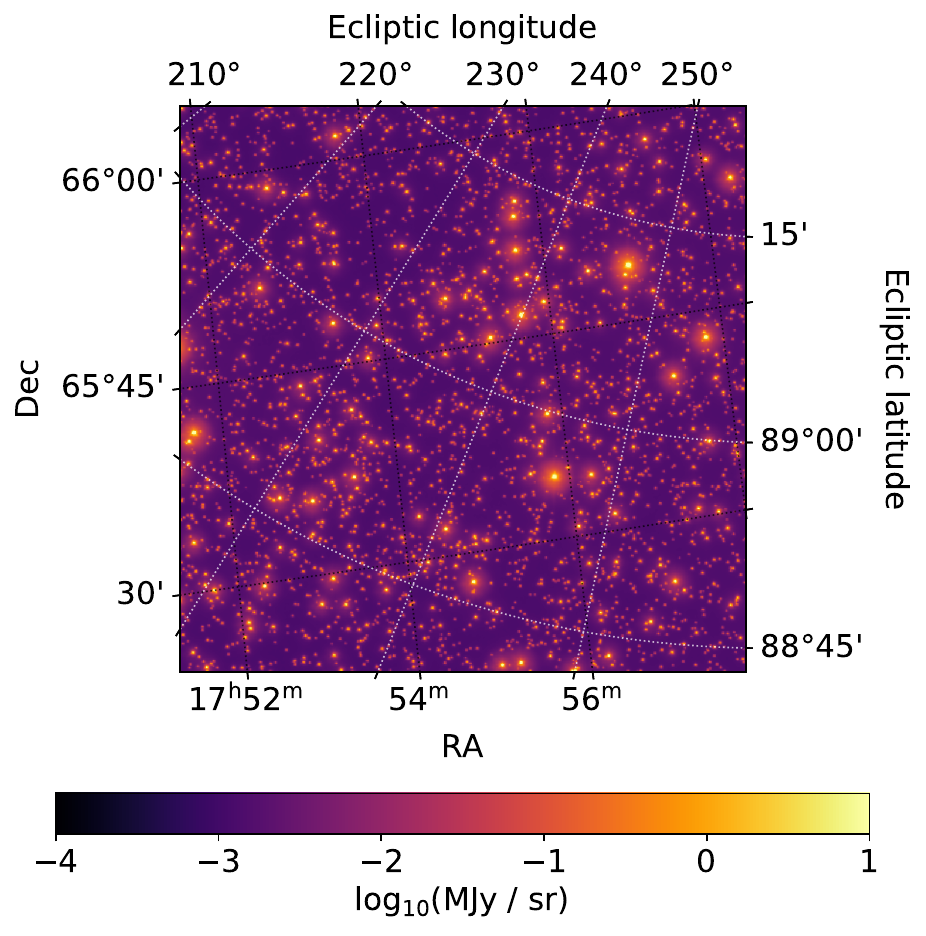}
    \caption{Thumbnail of the star component for band~1 for the pointing shown in \figref{fig:exampleImages}. Notice the extended-PSF halos. This image is produced by convolving the star catalog with the core~PSF and separately convolving with the extended~PSF~(\figref{fig:splitPSF}). The two convolved images are then summed to simulate the effect of the full~PSF.}
    \label{fig:starThumbnail}
\end{figure}

The PSF is essential for simulating and interpreting signals from unresolved sources such as stars and distant galaxies. In \figref{fig:starThumbnail}, we present a small area of a simulated image of starlight. Especially for bright stars, the PSF induces a halo that can be significant over an area comprising many detector pixels.

For computational efficiency in the \skysim, we maintain two types of PSF simulations, one of which focuses on fine-scale features close to the core and the other of which focuses on smoother, large-scale features in the extended component. We call the former the \emph{core~PSF} and the latter the \emph{extended~PSF}. The core~PSF is simulated with a resolution of~$\sim 1\arcsec$ and an extent of~$\sim 1'$, while the extended~PSF is simulated with a resolution of~$\sim 6\arcsec$ and an extent of~$\sim 3.5^\circ$. Although the platescale of our detectors is~6\parcs{15}, we are sensitive well below this scale. At the shortest optical wavelengths the PSFs are as small as~$\sim 3\arcsec$ 50\% enclosed energy fraction with a complex shape, and our forced photometry can be biased by the sub-pixel convolution. Additionally, photometry is sensitive to centroid positions at the level of~$\sim 0.1\arcsec$. For most of the bands, we simulate PSFs for a $3 \times 3$~grid of focal-plane positions. For the core~PSFs of bands~1 and~3, however, we expect greater variation and have simulated grids of $20 \times 20$~positions. For the short-wavelength corner of band~3 on the outside of the focal plane, we expect especially large wavefront variation, so we have oversampled this region with a smaller but denser $20 \times 20$~grid. 

Neither the core nor the extended~PSF is a complete characterization, though each may be individually sufficient for certain applications. For example, most of our forced photometry will depend entirely on the core. On the other hand, the extended PSF may be entirely sufficient for estimates of wide-angle contamination from bright sources such as planets. In other cases, such as the intensity mapping associated with the \spherex\ galaxy-formation investigation, both the core and the extended~PSF must be simulated in order to account for source contributions on all angular scales.

\begin{figure}
    \centering
    \includegraphics[width=0.9\textwidth]{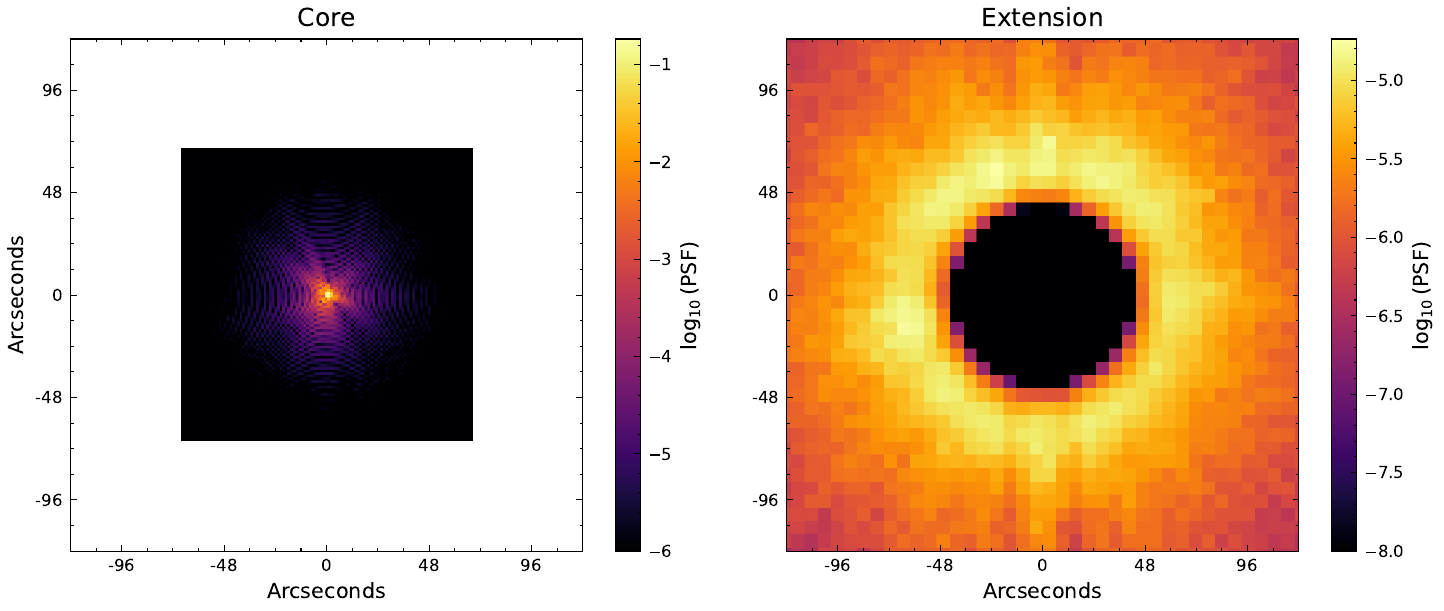}
    \caption{Split PSFs for the central detector pixel of band~3. The core~PSF is defined over a much smaller solid angle and is stored at finer resolution. The extended~PSF is stored over an area that is equal in size to the detector's field of view~($\sim 3.5\deg \times 3.5\deg$); for this figure, we have truncated to the central~$4\arcmin \times 4\arcmin$. The core-PSF simulations are tapered in their outer radii, and the extended-PSF simulations are tapered in their inner radii. The two are consistently normalized. The split PSFs can be convolved separately, and the resulting images can be summed to simulate the effect of the full PSF~(as in \figref{fig:starThumbnail}).}
    \label{fig:splitPSF}
\end{figure}

To handle both the core and the extended~PSF, we modify the two simulations to be consistent with each other. We impose a radial taper on the outer regions of the core~PSF, and we impose an inverted and complementary taper on the inner regions of the extended~PSF. An example of this type of split PSF is shown in \figref{fig:splitPSF}. We normalize the sum of the two PSFs but keep them separate. It is computationally efficient to restrict the spatial extent of the fine-resolution core~PSF and to restrict the resolution of the large-scale extended~PSF. As convolution is a linear operation, the two PSFs can be simulated separately. The resulting images can be summed to simulate the combination.

\begin{figure}
    \centering
    \includegraphics[width=0.85\textwidth]{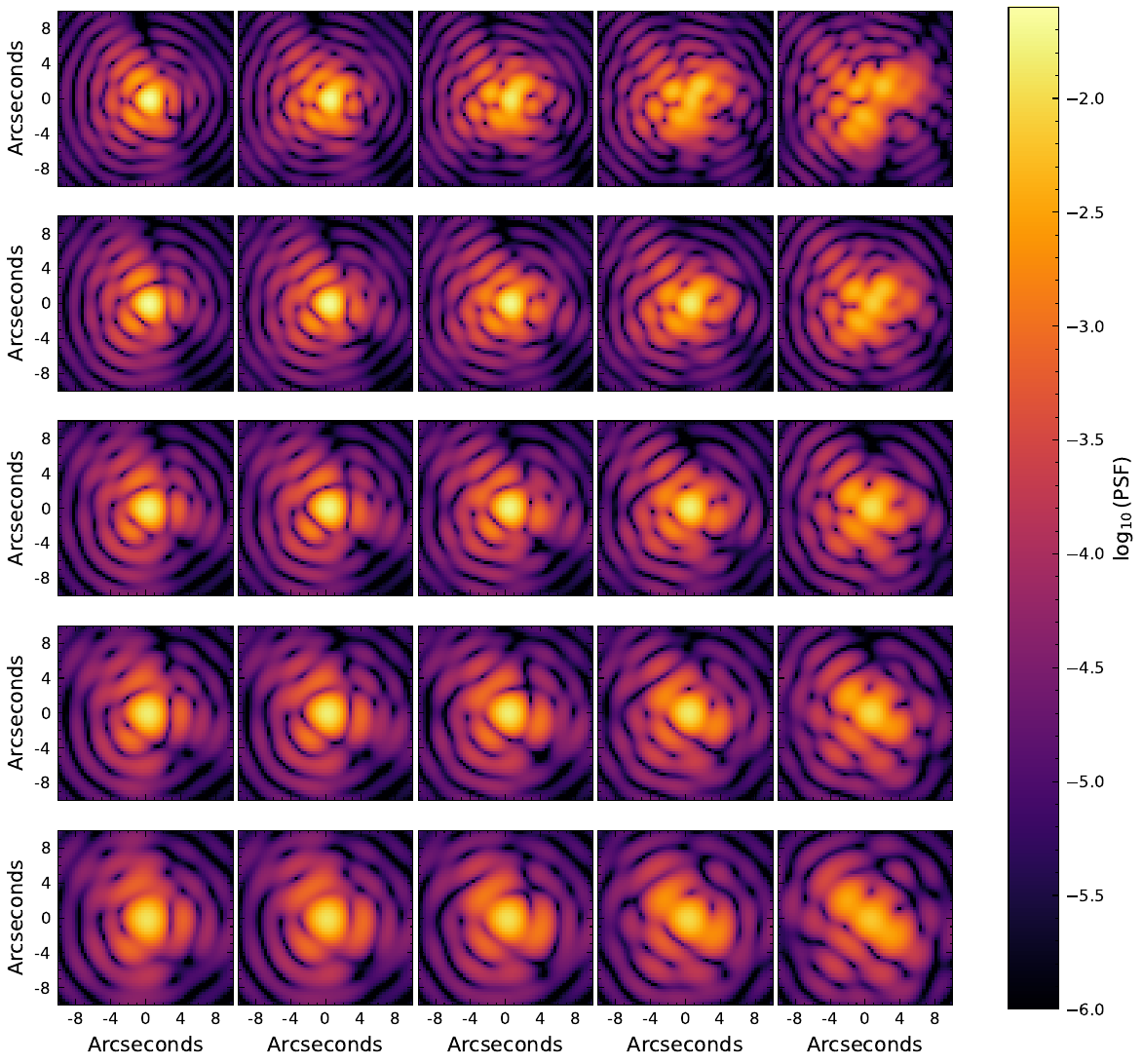}
    \caption{\label{fig:psfPanel}Panel of PSFs at evenly-spaced focal-plane positions on band~3. Notice that the PSF varies continuously in both directions across the band. In the downward direction, the PSFs generally widen due to an increasing wavelength; in the horizontal direction, the variation is mainly due to a de-focusing of the telescope optics. The underlying optical simulations provide PSFs at discrete locations, and the \skysim\ can interpolate among those locations.}
\end{figure}

Every \spherex\ detector pixel has its own true PSF, but it is computationally impractical to simulate and store all of them. Instead, we rely on the relatively sparse sampling that was described above. In some applications, it may be acceptable to ignore the variation in the PSFs as a function of focal-plane position~(which is also, as a result of the LVFs, entangled with the active wavelength). When the variation needs to be accounted for, we interpolate among the simulated PSFs, so there is effectively a smoothly-varying and unique PSF for every detector pixel. In \figref{fig:psfPanel}, we present a selection of position-dependent PSFs. In the vertical direction, much of the variation is due to LVF-induced wavelength changes. In both directions, there are variations in the telescope optics, so there are also PSF changes within each spectral channel.

The PSFs vary across the focal plane for two reasons: 1)~the detector wavelengths vary in the direction of the LVF gradient and 2)~the telescope optics creates different PSFs even along contours of constant wavelength. In effect, there is a different PSF for every single \spherex\ detector pixel, of which there are~$\sim 2.4 \times 10^7$. To avoid the computational expense of individually handling all of these PSFs, we have developed two methods of approximation. The two methods are similar in form but diverge most strongly at the step of PSF~convolution.

The faster method assumes that the PSFs are constant over large areas of each band; typically, we divide the band into a $3 \times 3$~grid and assign a single PSF to each block. Convolution is performed on each block, and the results are assembled to form a simulated image. We will refer to this method as \emph{discrete convolution}.

The slower but more realistic method also relies on discretized PSFs but constructs a \textit{blending} that simulates the effect of PSFs that are continuously varying across the focal plane. In each block of the band, we perform multiple convolutions corresponding to the nearby discretized PSFs acting on the extreme wavelengths of the chosen detector area. The results are combined with pixel-dependent weights that are based on spatial and spectral proximity. We will refer to this method as \textit{interpolated convolution}.

For a given pointing, each band is sensitive to light from a well-defined region of the sky, and the first step in both of our methods is to collect a catalog of sources that are located in or near the active sky region. This is done for each band separately, and different source types can be handled separately. For example, one can simulate stars for band~3 and then separately simulate galaxies for band~3. The results can be summed in a higher-level step. It is helpful to distinguish between interior sources whose centroids lie within the field of view, and exterior sources whose centroids lie outside the field of view. For discrete convolution, we consider only the interior sources. This choice is computationally advantageous, but it ignores the bright, exterior sources whose PSFs contribute non-negligibly to the exposure. For interpolated convolution, we consider both interior and exterior sources, where the latter are restricted to a \emph{buffer region} around the field of view. The size of the buffer region is set by the greatest extent of our PSF model. We perform convolution on the extended area and then crop the results to the field of view. There is a computational cost due to the increase in the number of sources and an increase in the area for convolution, but this approach ensures the inclusion of bright, nearby, exterior sources that contribute through the off-center portions of their PSFs.

We first form a \emph{high-resolution scene}, which is an oversampled representation of the exposure image. For interpolated convolution, we add a buffer region on the periphery of the image. When interested in the fine features associated with the core portions of the PSFs, typical oversampling factors are~5--10; this means that we simulate $5^2$ -- $10^2$~subpixels within each detector pixel~(and, when relevant, within the effective pixels in the buffer region). When interested in larger-scale features associated with the extended portions of the PSFs, oversampling may be unnecessary. The high-resolution scene also simulates LVF-induced wavelength variations on scales smaller than the true detectors.

Next, we aggregate source fluxes into the high-resolution scene.  This process relies on a choice for the effective wavelength at each high-resolution pixel. For discrete convolution, we choose the oversampled LVF wavelength. For interpolated convolution, this step is performed in smaller blocks, and two wavelength slices are used for each block. In each of those slices, the wavelength is taken to be uniform. After convolution, the wavelength slices will be blended with pixel-dependent weights in order to recover the LVF wavelengths.  To estimate the wavelength-dependent flux, we interpolate from the photometry included in the Reference Catalog. 

Although the high-resolution scene is oversampled, the pixel discretization is detectable as a random offset between the true centroids and the simulated centroids. The aggregation procedure forces the source centroids onto the high-resolution grid, and \spherex\ astrometry is sensitive to this effect at the level of~0\parcs{1}. Oversampling by more than a factor of~$\sim 10$ is computationally infeasible for applications requiring large numbers of simulated exposures, so the high-resolution pixels are typically much larger than our astrometric sensitivity. If these random offsets are ignored, simulations of forced photometry will show a bias due to a mismatch between the assumed and effective source position~\citep{Portillo_2020}. For \spherex\ PSF sizes and expected pointing jitter performance of 0\parcs8~RMS, this bias is $\sim$1\% for the shortest wavelengths at an oversampling factor of 5.  Oversampling at a factor 20 reduces bias to the 0.1\% level.  If further precision is required for a simulation application, the \skysim\ allows for the aggregated fluxes to be shared among neighboring pixels with divisions based on proximity. This comes at the cost of an additional spreading of flux, but it ensures that the sources are properly centered. This correction can be turned on or off depending on the application of interest.

\begin{figure}[ht!]
\plotone{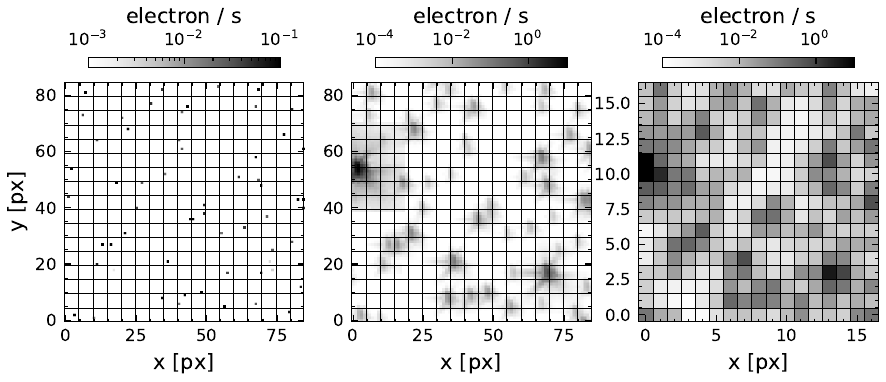}
\caption{An example of inserting compact sources with detector 2.  Left: source flux is introduced as delta functions in an oversampled high resolution image; Center: the high resolution image is convolved with the core PSF model; Right: finally, the image is rebinned to \spherex\ native resolution.  The grid lines show \spherex\ detector pixel edges.
\label{fig:core_psf_insertion}}
\end{figure}

Now we convolve the high-resolution scenes with our assumed PSFs. As described above, this is accomplished differently for the two convolution methods. Since information regarding where charge carriers were initiated in the material will be lost upon rebinning to the native {\spherex} resolution, we can convolve the high-resolution scene with an electron diffusion kernel as discussed in Section~\ref{sec:diffusion} before finally rebinning to recover the true detector pixel sizes. Figure~\ref{fig:core_psf_insertion} illustrates the source generation, PSF convolution and rebinning process.

For high-resolution extended signals~(\secref{sec:highresextsig}), the process is similar to that of compact sources~(\secref{sec:compactsources}). Because the input maps are given as wavelength slices, the PSF~convolution can proceed as described above for interpolated convolution. The detector is divided into blocks. In each block there are four nearby PSFs, and there are two extreme wavelengths. Convolved maps are made at both wavelengths and for all four PSFs, so there are eight maps that are blended to form detector signals. As in the case of compact sources, the blending is accomplished with pixel-dependent weights that are based on both spatial and spectral proximity.

\subsubsection{Bandpass}
\label{sec:bandpass}

The bandpass of \spherex\ is mainly determined by the linear variable filters mounted in close proximity to the detectors, as well as the dichroic beam splitter that separates the mid-infrared bands 4, 5, and 6 from the short-wave bands 1, 2, and 3. The model for the spectral response in the \skysim\ originally contained the nominal analytical model. Later, empirical model based on measurements during the instrument-level pre-launch integration and testing campaign is implemented as explained below.

For each pixel, the response, referred to as the ``bandpass,'' $T(\lambda)$, was measured \citep{Hui2024}.  The median of the curve, $\lambda_m$ (defined such that $\int_{-\infty}^{\lambda_m} T(\lambda) d\lambda = \int_{\lambda_m} ^{+\infty} T(\lambda) d\lambda$), was then calculated for each pixel based on the measured data. 

For simulation purposes, a smooth analytical function to model $\lambda_m(x, y; i)$ for the pixel $(x, y)$ and the detector band $i \in \{1, 2, 3, 4, 5, 6\}$ was sought. First, for each array, pixels with low quantum efficiency or high dark currents were masked as bad pixels. Additionally, pixels with negative or very small equivalent width ($\int_{-\infty}^{+\infty} T(\lambda)/\mathrm{max}\{T(\lambda)\} d\lambda < 15 \,\mathrm{nm}$) were masked, primarily to exclude pixels with extremely low signals due to the dichroic beam splitter. Then, the best-fit two-dimensional Chebyshev polynomials were determined for each array to model $\lambda_m(x, y; i)$. A second-order polynomial in both axes ($\tilde{\lambda}_{m, 2, 2}(x, y; i)$) was found to be sufficiently accurate (with residuals below 1\% of the spectral resolution for $>99\%$ of the pixels), and higher-order fits did not significantly reduce the residuals.

We then grouped pixels based on $\lambda_m(x, y; i)$ (e.g., $\pm 0.25\,\mathrm{nm}$ range). Typically, a few thousand pixels (1 to 2 pixel-thick curves) share the same $\lambda_m$ values. For each of these $\lambda_m$, the bandpasses ($T(\lambda)$) of the corresponding pixels were selected, normalized, stacked, and the representative (median) curve was computed. This curve is then linearly interpolated to a finer grid (a few nm) for simpler simulation calculations. Any negative values in this curve were replaced with 0. The final curves for each $\lambda_m$ were collected and saved into a parquet file to serve as the bandpass template. The template curves typically have an RMS error of $\sim 0.001$ (based on $T_\mathrm{simulator}(\lambda) - T(\lambda)$, where each $T$ is normalized). For all bands except band 4, the 99.9th percentile error is $\lesssim 0.002$. For band 4, residuals are slightly larger due to the dichroic edge effects, with the median and 99.9th percentile errors of 0.006 and 0.007, respectively. As a result, for each band, approximately 800 to 3,000 unique curves were generated and stored. Unique integer IDs were assigned to each template curve, increasing monotonically with the corresponding $\lambda_m$. The exact structure of the templates may be updated in the future. 

An auxiliary file of two-dimensional integers for each band, mapping pixel locations to template IDs, was required. This was generated by identifying the ID of the closest $\lambda_m$ in the template to $\tilde{\lambda}_{m, 2, 2}(x, y; i)$ for each pixel. For the outer 4-pixel-thick reference pixel area, where the full bandpass is unavailable, this 2-D integer map of 2040 by 2040 pixels was padded based on the 3-pixel median of the outer edges.

Given a pixel location, the bandpass center is approximately retrieved using the analytical function $\tilde{\lambda}_{m, 2, 2}(x, y; i)$. The full bandpass function is obtained by referencing the template file, where the template ID is available in the auxiliary file.

See Figures~\ref{fig:SWIRbandpasses} and~\ref{fig:MWIRbandpasses} for schematics of the bandpass models. 

\subsubsection{Optical Ghosts}

In pre-flight laboratory data, we have identified two sources of optical ghosts which transfer power to far outside the PSF core and have developed models to simulate both.  
These effects act on the images of sources both inside and outside of the physical detector, and require source catalogs well beyond the nominal edge of the images. 

The first effect is caused by sources whose images fall on the housing that holds the LVF filters above the detector arrays, which we term \textit{frame edge ghosts}.  This is analogous to the ``Dragon's Breath'' scattering observed in \textit{JWST} data \citep{JWSTproblems}, but with a different, \spherex-specific geometry.  The geometry and tests used to characterize this effect are fully described in \citet{Nguyen2024}.  For the \skysim, we implement a model that takes as input a catalog of sources at angles $37^{\prime \prime} < \theta < 285^{\prime \prime}$ beyond the edge of each array, corresponding to a range of angles which generated the effect in lab measurements.  A model image is generated by modeling a spike of light based on the measured coupling between the incident flux and ghost brightness, shown as the red lines in Fig.~7 (spike amplitude), Fig.~8 (spike length) of \citet{Nguyen2024}.  We chose to model the width of the spike as a Gaussian with fixed $\sigma = 1.2$ pixels for all angles of incidence.  The spikes are placed at the source position with the long axis orthogonal to the edge of the detector array, as light would scatter if reflected from the frame edge.

The second optical ghost effect is caused by 
internal reflection in the dichroic beam splitter (DBS).
This reflection path produces an out-of-focus, approximately circular image of the source offset $\sim$24 arcmin from the source. 
The integrated amplitude is as high as $\sim$1\% in the narrow wavelength range of the beamsplitter in Bands 3 and 4 and is otherwise in the $10^{-4}$ to $10^{-5}$ range for most of bands 4-6.  Absorption in the DBS substrate prevents the DBS ghost for bands 1, 2, and most of 3.

Optical modeling has led to the following approximate formula for the DBS ghost location:
\begin{equation}
  x_{g,i} = x_{s,i} + \alpha_{ij} x_{s,j} + o_{i}
\end{equation}
with the labels $x_{g}$ for ghost image position, $x_{s}$ for source image
position, `o' for offset, and the indices $i$ and $j$ running over
the $x$ and $y$ positions.   
The parameters of this transformation for each band are given in Table
\ref{tab:DBSghosts}.  The coupling amplitudes that convert source flux to surface brightness on the detector vary with wavelength across each of the four bands, but are typically $\sim 10^{-4}$ integrating over the ghost image. In all cases, the size of the ghost has $r_{g}=51$ pixels centered on $x_{g}$, with variable coupling amplitudes that are based on the 
optical modeling.  
The geometry of the optical chain imposes a region of interest for sources that can cause ghosts, which in the case of the DBS ghosts falls both on the array and in the area around it.  In band 3 this region subtends an area of 2.6 square degrees concentrated at the long wavelength side of the detector array, and in bands $4{-}6$ an area of 13.9 square degrees again skewed slightly to the long wavelength side of the detector array.

\begin{table}[ht]
  \centering
  \caption{Dichroic Beam Splitter Ghost Model Parameters  \label{tab:DBSghosts}}
  \begin{tabular}{l|cccccc}
    \hline
Band & $\alpha_{xx}$ & $\alpha_{xy}$ & $o_{x}$ & $\alpha_{yx}$ & $\alpha_{yy}$ & $o_{y}$ 
\\ \hline
3 & $-0.0051$ & $0.0013$ & $-7.4$ & $0$ &  $0$ & $225$  \\ 
4 & $-0.0040$ & $-0.0013$ & $-8.9$ & $0$ & $0$ & $225$  \\ 
5 & $-0.0042$ & $0.0005$ & $0.1$ & $0$ & $0$ & $225$ \\ 
6 & $-0.0043$ & $0.0013$ & $9.4$ & $0$ & $0$ & $225$  \\ \hline 
  \end{tabular}
\end{table}

\subsection{Signal Transduction}

The \skysim\ converts between astrophysical surface brightness in \MJysr and photocurrent in \eps produced in the detector with a constant factor for each of the six detectors:  
\begin{equation}
    g = \frac{A \Omega}{Rh} \eta_\mathrm{opt}
\end{equation}
where $A$ is the telescope collecting area (assumed to be a 10~cm radius circular aperture), $\Omega$ is the solid angle of the pixel projected onto the sky, $R$ is the spectral resolution $\Delta \lambda / \lambda$, $\eta_\mathrm{opt}$ is the end-to-end band-mean optical efficiency, and $h$ is Planck's constant.    See Table~\ref{tab:spherex_bands} for values of the parameters for each of the six detectors.

The end-to-end band-mean optical efficiency $\eta_\mathrm{opt}$ is the ratio of power arriving at the detector to the power arriving at the aperture that is the product of the efficiency of all optical elements measured or modeled at component levels: the mirrors, the dichroic beam splitter, the linear variable filter, and the detector itself.  The values, between roughly 0.5 and 0.8 are listed in Table~\ref{tab:spherex_bands}.

Relative pixel-to-pixel variation in this conversion factor is encapsulated in the normalization of the bandpass (which captures efficiency variation as a function of wavelength due to the LVF and dichroic beam splitter) and in the relative gain map.

The relative gain map is derived from laboratory optical testing conducted at the focal plane level. As part of the spectral calibration test campaign, the detector was uniformly illuminated using multiple broadband diffuse sources delivered through an integrating sphere and Winston cone setup, as described by \cite{Hui2024}.  
This data product is normalized to achieve an band mean response of 1.0; it is then multiplied by the band-mean end-to-end efficiency to obtain a pixel-by-pixel end-to-end efficiency. 

\subsection{Readout Simulation and Flag Generation}
\spherex\ detector arrays use an on-board sample up-the-ramp (SUR) algorithm to measure pixel photocurrents. A slight variation of the original baseline model \citep{Zemcov2016} is implemented to the onboard software, and we briefly summarize the algorithm here.

Each non-destructive read is performed every $\approx 1.5 \,\mathrm{s}$ for $N_\mathrm{samp}$ times, resulting in a total exposure time of $\approx 112 \,\mathrm{s}$ in case of $N_\mathrm{samp}=73$, the value used at the time of launch and currently adopted in the \skysim. The algorithm calculates the linear least-squares slope estimator for the photocurrent using all valid frames for each pixel.  The algorithm also monitors the incoming frames for \textit{overflow} and \textit{transient} conditions and halts the frame accumulation for the affected pixel if one or both of the conditions are found.  The overflow threshold is set approximately half way between the reset and saturation values and is designed to keep the nonlinearity of the detector signal under $\sim 10\%$.  The transient detection is intended to reduce the impact of cosmic rays on the measured photocurrents.  In detail, each frame is compared to the 4 prior frames (so that the  estimator for the pixel value, the accumulated electron number, at the $i$-th frame $y_{i}$ is $\hat{y}_{i} = y_{i-1} + (y_{i-2} - y_{i-4})/2$). A transient flag is triggered within the onboard software when $y_i$ deviates from $\hat{y}_{i}$ by a programmable magnitude.
In the \skysim, the full SUR cube can be simulated first for convenience. Then the flags and the final slopes (total electron count divided by the exposure time) are generated at the last stage to mimic the onboard software. 


The \skysim\ does not explicitly simulate a stochastic population of energy deposition in pixels due to space radiation in LEO, rather it extrapolates transient rates from WISE data and flags a corresponding number of random pixels in each exposure.  If the survey plan indicates that the spacecraft is located in the South Atlantic Anomaly  during an exposure, the number of  randomly-flagged pixels is increased by a factor of 100. 

In the current version of the \skysim, the algorithm classifies a new SUR value as transient if it exceeds approximately the 5-sigma bound (accounting for both read noise and Poisson noise), using empirically determined coefficients based on lab measurements for each detector, as of the current version of the \skysim.

\subsection{Detector Effects}

Following conversion to electrical signals, a variety of effects in the detector substrate and amplification chain impact the data that must be simulated.  In general, these can be simulated as a series of additive terms that are modeled using information from pre-flight testing.


\subsubsection{Dark Current}

\begin{figure}[ht!]
\plotone{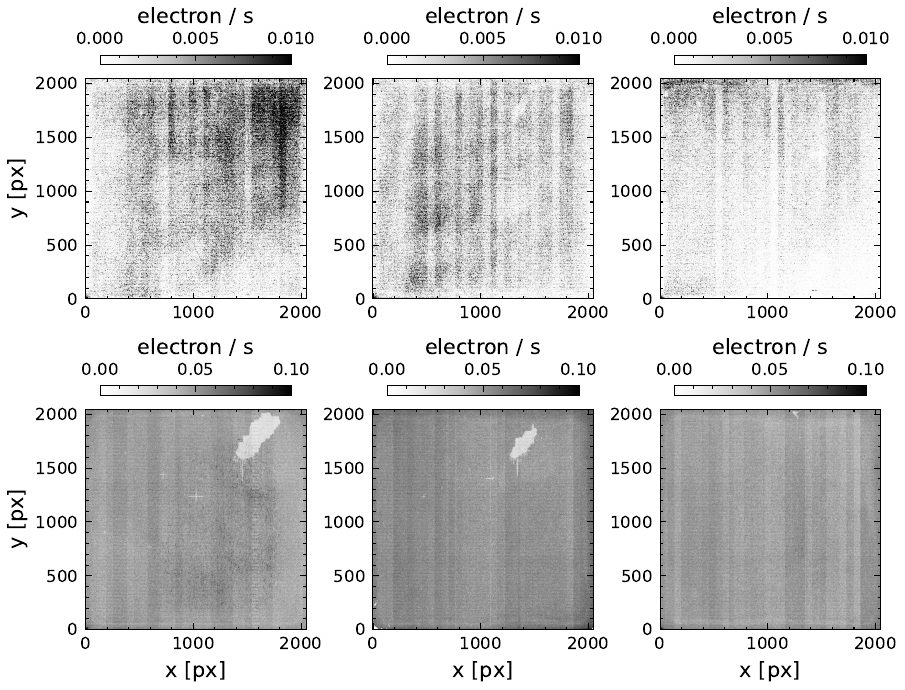}
\caption{Dark current templates used in the \skysim.  The top row are the short-wave detectors 1,2,3 and the bottom row are the mid-wave detectors 4,5,6.  These templates are used to generate random draws of dark current for each simulated exposure.
\label{fig:darkcurrent}}
\end{figure}

Charge carriers are produced in the detector substrate even in the absence of external illumination, an effect called \textit{dark current}.  The \skysim\ ingests laboratory measurements of the dark current generated as described by  
\citet{Nguyen2025}. 
To reduce noise from propagating into realizations of dark current in the simulated exposures, the dark current measurements have additional filters applied to create the simulation templates.  First, a 4-pixel box car median filter is applied to the measured dark current.  Second, medians below 10$^{-4}$ \eps\ are set to this value to prevent the \skysim\ from producing negative dark current values.  The resulting dark current templates are shown in Fig.~\ref{fig:darkcurrent}.  The templates provide a pixel-by-pixel mean current; for each realization a random Poisson-distributed number is drawn for each pixel and the resulting dark current is applied as an additive term to the simulated image.

\subsubsection{Noise Simulation}

\spherex\ data are subject to three primary noise sources: shot noise due to the random arrival of photons, random noise in the detector, and random noise in the amplification chain.  In the \skysim\ mode where every ramp sample is generated, the shot noise term is drawn from a Poisson distribution whose mean is the square root of the number of electrons in a pixel.  This is randomly generated for each exposure as an additive term.  

The detector and amplification noise are combined together into a single term called \textit{read noise} that we model using measurements of the as-built system.  To accurately simulate the properties of this noise source we model it using a per-pixel white noise term and a frequency-dependent $1/f$-spectrum per-channel noise term  \citep{Heaton2023,Nguyen2025}.  Each H2RG has 32 readout channels that each span 64 pixel columns.  These all have slightly different noise properties compared to the mean of the entire array, which imprints a vertical noise structure in addition to the pixel white noise.
The \skysim\ can generate readout noise in two ways. The first method implements the theoretical model of \citet{Heaton2023} to generate $1/f$ noise in every SUR sample, including a component that is correlated between readout channels.
The frames are passed through the SUR algorithm to obtain the photocurrent.
However, this method is computationally intensive and for many applications the added modeling fidelity is unnecessary.

\begin{figure}[ht!]
\plotone{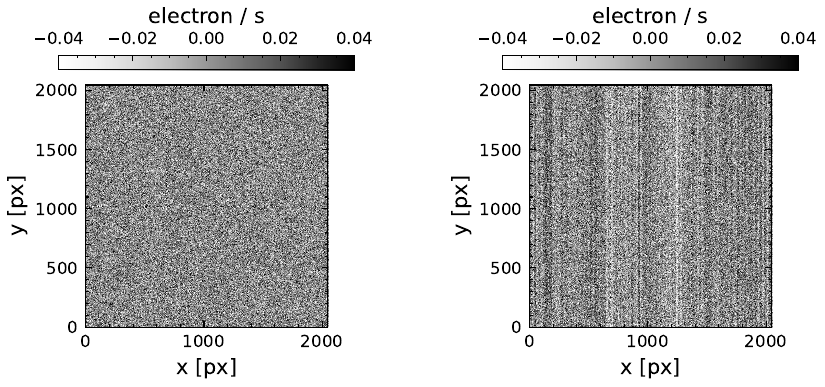}
\caption{An example readout noise realization for detector 2. Left: per-pixel noise; Right per-pixel plus correlated readout channel noise.  Noise is realized for each simulated exposure and summed into the output image as an additive term.
\label{fig:readnoise}}
\end{figure}

A second, computationally faster method was developed to  simulate the noise in a more approximate manner.  In this approach, we measure the mean noise and variance of every pixel and the mean pixel-to-pixel variance in each of the 32 readout channels using laboratory data read out as in flight exposures.
To simulate the per-pixel noise component, for each pixel, we draw a random number from a Gaussian distribution whose mean is the per-pixel mean noise, and whose sigma is the per-pixel variance.
Next, a per-channel noise component is generated in a similar fashion, but with all pixels in a readout channel drawing from a Gaussian distribution whose mean and variance matches that channel's noise offset and variance.
The per-pixel and per-channel noise realizations are then summed together, which examples shown as Fig.~\ref{fig:readnoise}.   Alternating columns within each readout channel are modeled with independent random number draws to model odd/even column noise.
The photon noise term is scaled by a multiplicative factor of $\sqrt{6/5}$ to account for the statistics of the fitted ramp slope in the presence of shot noise \citep{Garnett1993} and added to the read noise in quadrature to model the full noise in each pixel.

\subsubsection{Charge Diffusion}
\label{sec:diffusion}

We model charge diffusion as a two-dimensional Gaussian kernel based on \cite{Mosby_2020} with a standard deviation spreading parameter of approximately 0.1 pixels, derived from {\spherex}  
array parameters. This kernel is then convolved with the high-resolution scene, simulating diffusion of photo-generated charges before rebinning to the native {\spherex} 
resolution.  The resulting effect is a slight broadening of the PSF spread, particularly for point sources near the edge of a pixel. Since the PSFs used prior to this step are based on an optical model and not measurements, this charge diffusion simulation is not a duplicate.

\subsubsection{Non-Linearity}

The capacitance of a non-destructive photosensitive diode changes as the diode accumulates charge, resulting in a decreasing voltage accumulation rate for a constant flux which drops to zero rapidly at saturation.  We model \spherex\ non-linearity and saturation after the empirical model in \cite{Biesiadzinski2011}, with parameters based upon lab measurements of the detector full well ($b=186$~ke$^-$ for bands 1-3, and 129~ke$^-$ for bands 4-6) and hard-coded parameters (capacitance linearity parameter $a=1$ and the exponentially decaying dark-current time-constant $\tau = 100\,\mathrm{s}$ in Appendix A of \citealt{Biesiadzinski2011}). Because the dark current is added to the simulated SUR images separately, no additional exponential dark current is applied. The saturation roll-off can be seen in the SUR read-out simulation.

\subsubsection{Electrical Crosstalk}

The H2RG multiplexer layer implements bias and clock voltages that are common between the 32 detector outputs.  Pixels with large fluence levels in a given integration can electrically crosstalk via the detector layer, multiplexer layer, and amplification circuit, the cumulative effect of which causes pixels in the modulo-64 pixel positions from the high-fluence pixel to register a different current than it would based purely on the incident photo-flux.  For the \skysim, a model for the electrical crosstalk is developed using laboratory measurements of spatially structured sources.  Several models of differing complexity are available, depending on the needs of the particular simulation.  Electrical crosstalk always occurs in the fast electrical read direction, which for \spherex\ is designed to be orthogonal to the direction of spectral variation.  The electrical crosstalk model is applied to all pixels in an exposure, though only pixels with $i > 500$ \eps\ demonstrate an appreciable effect, and all but the brightest pixels source crosstalk images that are beneath the noise level in a simulated exposure.

The simplest model uses empirical measurements of the crosstalk amplitude between each channel and all other channels in the array, with those at separations $> 16$ channels being drawn from a statistical model.  The dominant crosstalk occurs between channels separated by $\pm 2$ output channels, that is, a source in channel $n$ has largest crosstalk with channels $n \pm2$ at the level of $\sim 10^{-4}$.  The $n \pm 1$ channels have couplings that are typically $< 1/2$ this amplitude.  Crosstalk in channels $\pm 3$ away and further drop roughly exponentially with distance from the incident source.  The minimum crosstalk amplitude is $\sim 10^{-7}$ for the output combination separated by 31 electrical channels.

A better approximation to the measurements is given by a model where we impose a variable gain between the incident photocurrent and the crosstalk amplitude, which arises due to full-well saturation in the source pixel during the exposure.  The gain term looks like a logistic function with a coupling amplitude rolloff beginning near 100 \eps, a $-3 $dB point at 2000 \eps, and negligible crosstalk amplitude by $10^4$ \eps.  This causes model images of very bright sources to have a characteristic annular shape, since the largest crosstalk coupling is occurring for pixels with photocurrents of several thousand \eps.  

Inspection of laboratory crosstalk measurements shows there are actually two crosstalk patterns, one which is proportional to the source signal, and the other resembling a derivative of the source signal in the readout direction \citep{Finger2008}.  The most accurate model for these as-measured data updates the crosstalk coupling matrix to permit both positive and negative coupling terms, with coupling to a bright pixel possible for the neighboring pixels along the read direction.  In a given crosstalking channel, the characteristic shape of the crosstalk is a dipole of neighboring pixels in either the positive-negative or negative-positive orientation, with roughly equal amplitudes between the two terms.  This makes the images of the crosstalk effect most sensitive to the derivative of the intensity distribution, since sharp discontinuities between the brightnesses of neighboring pixels cause the largest dipole image in the crosstalking channels.  The corollary is that diffuse signals with slow derivatives, even if bright, tend to cancel and minimize the impact of the crosstalk.  The amplitudes of the crosstalk in both the positive and negative senses are determined from laboratory measurements and implemented in the model as a three-dimensional gain matrix, with the third dimension being the crosstalk amplitudes in the $\{-1,0,1\}$ pixel shift directions.  

\subsubsection{Inter-pixel Capacitance}

It is well known that adjacent detector array diodes with differing bias voltages will be pulled towards an intermediate voltage due to inter-pixel capacitance (IPC) (\cite{Moore_2004}).  In the \skysim, the strongest effect of IPC is represented by a $3 \times 3$ signal independent convolution kernel matrix modeled after \cite{Mosby_2020}.  In this matrix, $2.0\%$ of the central pixel's signal is detected in each of its nearest neighbors and $0.2\%$ in each of its corner neighbors, leaving $91.2\%$ of a pixel's signal in the central pixel.  After all other charge-contributing effects are considered, this kernel is convolved with the resulting detector image, thus simulating IPC.

\subsubsection{Image Persistence}

After exposure to illumination, the detector diodes exhibit a decayed form of the illuminated image in subsequent exposures, an effect known as image persistence.  In the \skysim, we implement the inverse-time model of this phenomenon described in \cite{Fazar2025} with detector-specific parameters, the magnitude of which depends upon the total number of charge carriers generated in a pixel. 
After a selection of sky images are generated, the images can be run through the image persistence module in the \skysim, which looks back at the previous 500 exposures and calculates the amount of decayed image expected to be present in the current exposure based upon the fluence in each prior exposure and the time elapsed since the reset following that exposure.  These predicted persistence currents are then added together to form a persistence image, which is then added to the current exposure, simulating the persistent signal.

\section{Results and Sensitivity Predictions}
\label{sec:results}
The \skysim\ and its thorough modeling of the as-built instrument and expected sky signals enable an end-to-end point source sensitivity estimate for \spherex.    To perform this calculation, we distributed 768 artificial compact sources with zero flux over the sky on an evenly spaced grid on the pixel centers of a Healpix nside=4 tesselation of the sky.  

\begin{figure}[ht!]
\plotone{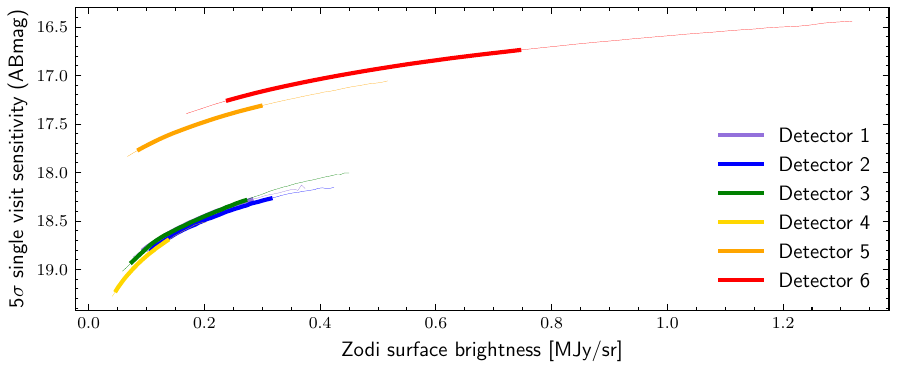}
\caption{Predicted \spherex\ 5$\sigma$ point source sensitivity in a single visit as a function of the local background surface brightness, which when dominated by zodi, depends on time of year and line of sight direction.  The domain of the curves shows the range of zodi background expected for each detector during the two-year \spherex\ survey, with the thicker line spanning the 5th to 95th percentiles. 
\label{fig:sensitivity_vs_zodi}}
\end{figure}

\begin{figure}[ht!]
\plotone{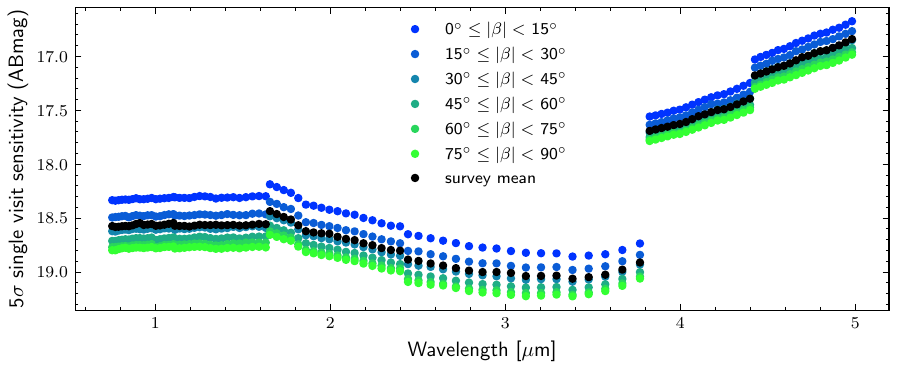}
\caption{\spherex\ 5$\sigma$ point source sensitivity in a single visit (with the pre-launch integration time 112.5~s) vs. wavelength as estimated by the \skysim; colors represent bins of ecliptic latitude $\beta$.
\label{fig:sensitivity}}
\end{figure}

Sky signal models of zodiacal light, and the instrument noise performance provided inputs into the QuickCatalog mode of the \skysim.   The full mission observation sequence from the survey planning software determines the temporal ordering of the samples of the spectrum of each of these sources, and the sky background at each time.    QuickCatalog created 10,000 noise realizations of read noise, dark current, photon noise from the background using the approximation that bypasses the full SUR. Confusion is not included.  The distribution of these realizations provide a 5$\sigma$ point source sensitivity, which is presented in Figs.~\ref{fig:sensitivity_vs_zodi} and \ref{fig:sensitivity}).  As expected, the sensitivity floor is set by the photon noise of the zodi background.  Figure~\ref{fig:sensitivity_vs_zodi} shows this dependence explicitly, for the range of zodi brightnesses expected during the course of the two-year mission.  The zodi brightness varies with both time and sky position and coupled with the survey strategy leads to an overall source flux uncertainty shown in \figref{fig:sensitivity}, which shows a clear dependence on ecliptic latitude.

The sensitivities shown in \figref{fig:sensitivity} are estimates for a single visit observation of a source in each spectral channel.  This corresponds to the uncertainty that a user of the \spherex\ data can expect when measuring photometry of a source in a single spectral image from the flight data.  The sensitivities expressed in previous publications \citep{Korngut2018} estimate the final sensitivity of \spherex's photometric measurements including averaging multiple measurements within each spectral channel over two years of the all sky and deep surveys.  In the limit where the noise floor is set by the zodi shot noise and every photometric measurement is successfully achieved, the all-sky survey sensitivity will be exactly a factor 2 deeper due to the four visits per spectral channel, while the deep survey in the 100 square degree patches near the NEP and SEP will be a factor of 10--20 times deeper than the single-visit sensitivities.  The full-survey sensitivity of \spherex's measurement of any particular source is not exactly predictable as photometric measurements contaminated by effects such as optical ghosting, crosstalk, or cosmic ray transients (affecting several percent of photometric measurements) will be flagged and discarded.

\section{Conclusion and Future Plans}
\label{sec:future}

We conclude that developing the \skysim\ software early in the mission life cycle (Phase B -- Preliminary Design Phase) of \spherex\ was a worthwhile investment.
A centralized and flexible facility to generate reliable data simulations, with both a high fidelity sky model and a high fidelity instrument model, has been critical in the development of \spherex, and is planned to continue to play a key role in the science analysis effort.  At time of writing, the simulator has been used for a variety of testing and validation activities.

One of the major activities simulated data has been deployed against is the development of the \spherex\ Level 1--3 data reduction pipeline.  The \skysim\ provided test data sets to create prototype code for pipeline tasks such as flat-field estimation, astrometric calibration, PSF reconstruction, earth-satellite detection and masking, saturated pixel masking, photometric calibration, and so on.  In addition, to test the Level 1-3 pipeline at scale, an entire year of simulated spectral image data -- 1,239,540 exposures in all -- was generated.  This simulation used 13000 node-hours on the stampede3 and lonestar6 systems at the Texas Advanced Computing Center.  The resulting simulated images were ingested and processed by the \spherex\ Science Data Center’s pipeline, allowing science center requirements to be verified at scale and for the team to gain experience with the science processing pipeline ahead of launch.

The simulator has been used to assess the impact of specific detector behavior on flight data, i.e. persistence current as described in \citet{Fazar2025}.  Many of the instrumental effects discussed in Section \ref{sec:instrument} have been studied with the \skysim, and will continue to be assessed using it.  These studies were enabled by developing the science simulation software in tandem with developing the instrument.

The \skysim\ has been central in prototyping and testing the Level 4 science pipelines. Outcomes include software to filter and mosaic tens of thousands of exposures, testing redshift-fitting procedures, prototype ices column depth fitting software, and to prototype systematic error evaluation code throughout the \spherex\ science themes.  

Following the March 11, 2025 launch of \spherex, the \skysim\ software will be updated in a number of ways to enhance its utility for the science data analysis phase of the mission.  First, the in-orbit checkout (IOC) phase included an optimization of the overall survey performance (sensitivity and voxel completeness) by configuring the instrument and the survey plan based on the spacecraft and instrument performance on-orbit. In particular, settings of the readout electronics, such as the exposure time and on-board sample-up-the-ramp flagging, will be adjusted.   These parameters will be updated within the \skysim\ to match the instrument settings used during the science survey.  

Next, as the \spherex\ science survey proceeds and our understanding of the observations improves the instrument performance parameters will become better characterized over time.  This includes improvements in our understanding of instrumental crosstalk, image persistence, optical field distortion, the point-spread function (both the core and extended), the frame edge ghost effect, the detector flat-field response and dark current.  The parameterization of these effects will be updated in the instrument model of the \skysim.

Finally, as our astrophysical understanding improves new sky signal models will be implemented to match phenomena observed on-orbit.  High fidelity simulations from the \skysim\ helped the team to identify unexpected features of the data.  This includes airglow effects that are not yet simulated; for example the \spherex\ first light press release images show excess signal at 1.083~\micron, which corresponds to a well-known geocoronal feature seen in the infrared channel on Wide-Field Camera 3 on the Hubble Space Telescope and attributed to HeI emission \citep{firstlightpress,Brammer2014}.   This emission will be included in the model of the diffuse background in simulated \spherex\ exposures in a future version of the \skysim.  Other effects will be included as they are isolated and modeled.

The updated \skysim\ will play a key role in the \spherex\ science data analysis.  The \skysim's ability to forward-model observational and instrument systematic errors will enable quantification of their impact on all science results.  For the EBL analysis, this will involve noise propagation and transfer function estimation for the power spectrum analysis.  The Cosmology analysis will use the \skysim's capability to inject simulated sources of known flux into real \spherex\ images to evaluate the selection functions and systematic errors.  These studies using the \skysim\ will be part of future \spherex\ science results papers as well as papers focused on specific systematic errors.  We plan to make public the future on-orbit version of the \skysim\ to support community activities.

\appendix
\section{\spherex\ bandpasses and spectral channels}
\begin{figure}
    \centering
    \includegraphics[width=0.9\linewidth]{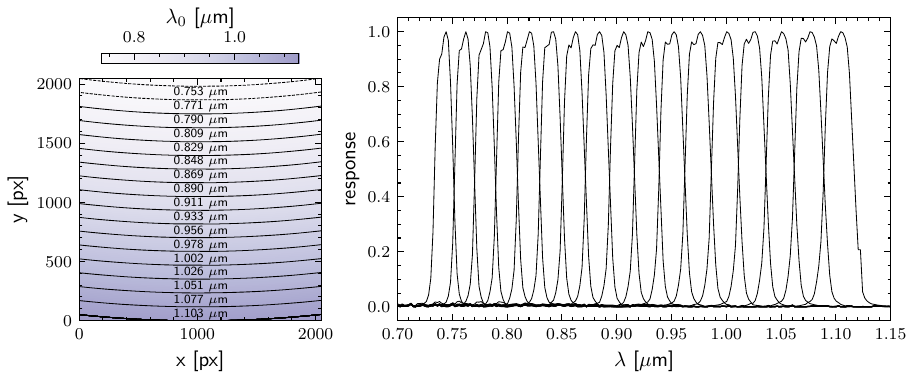}
    \includegraphics[width=0.9\linewidth]{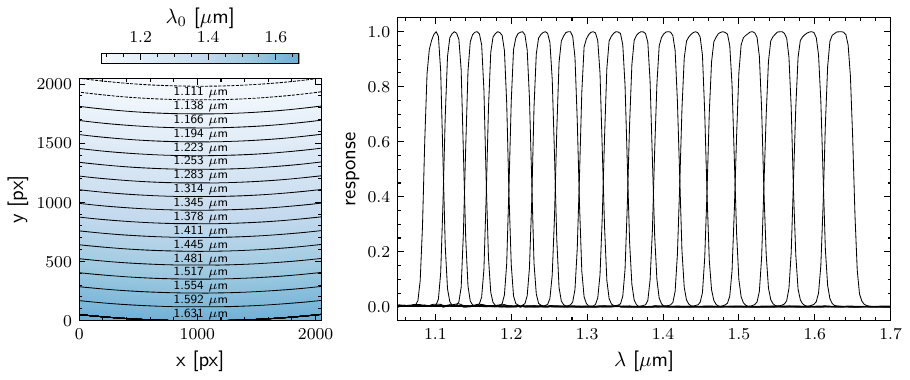}
    \includegraphics[width=0.9\linewidth]{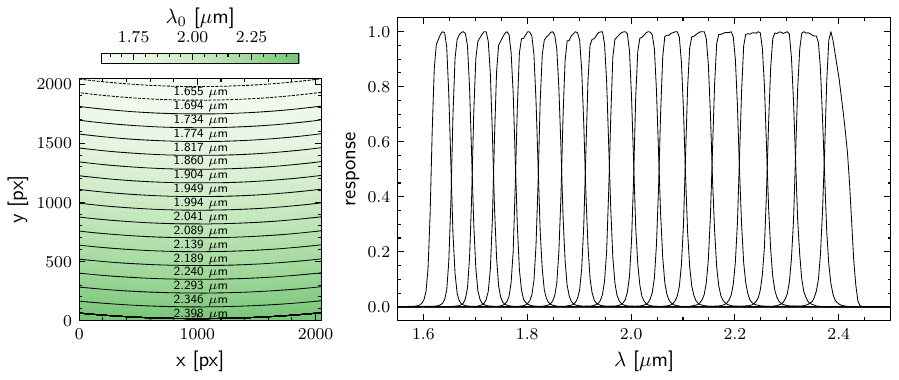}
    \caption{The \skysim\ bandpass model for the short-wave detectors 1,2, and 3 (top, middle, and bottom rows respectively).  Left column: the band center wavelength $\lambda_0$ as a function of position on the array with boundaries of the spectral channels indicated and labeled with the mean wavelength.  Right column: bandpasses response functions (normalized to peak response) for pixel locations in the center of each spectral channel.  The slight curvature is an effect known as ``smile" that is due to the linear variable filter manufacturing process.  The region of the detector outside of any spectral channel will also be present in the spectral image but is not included in the spectral coverage of the survey.}
    \label{fig:SWIRbandpasses}
\end{figure}

\begin{figure}
    \centering
    \includegraphics[width=0.9\linewidth]{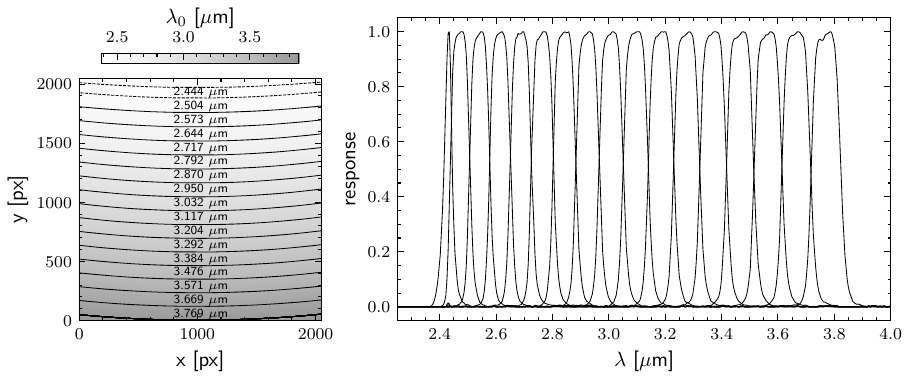}
    \includegraphics[width=0.9\linewidth]{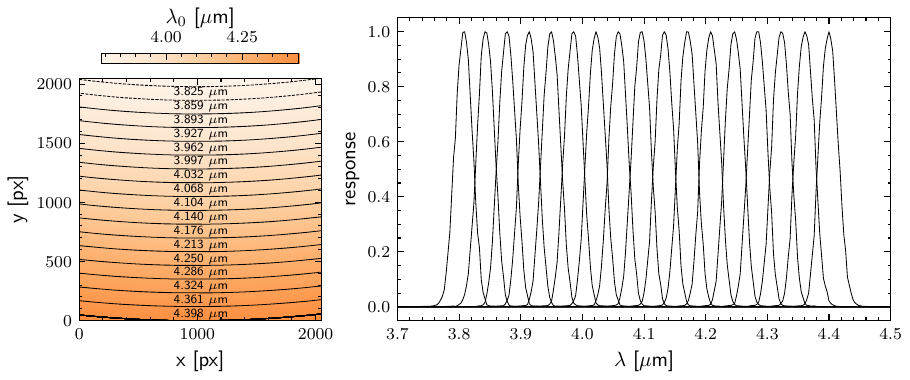}
    \includegraphics[width=0.9\linewidth]{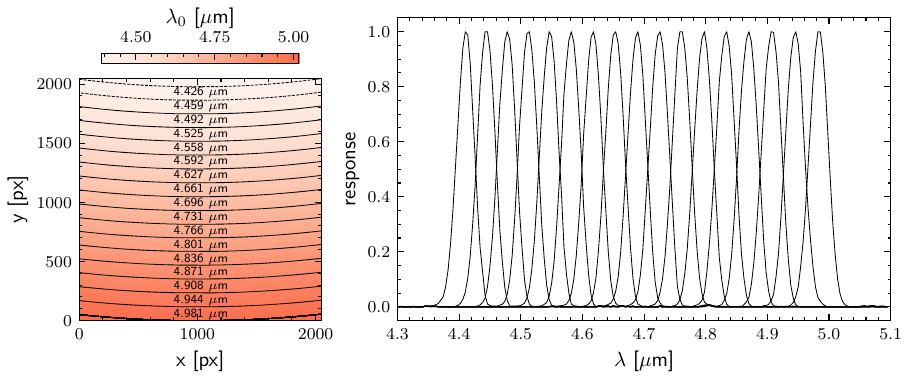}
    \caption{The \skysim\ bandpass models for the mid-wave detectors 4,5, and 6 (top, middle, and bottom rows respectively). Left column: the band center wavelength $\lambda_0$ as a function of position on the array with boundaries of the spectral channels indicated and labeled with the mean wavelength.  Right column: bandpasses response functions (normalized to peak response) for pixel locations in the center of each spectral channel. }
    \label{fig:MWIRbandpasses}
\end{figure}

Figure~\ref{fig:SWIRbandpasses} shows the bandpass models for \spherex\ detectors 1,2, and 3 and Fig.~\ref{fig:MWIRbandpasses}  shows the bandpass models for detectors 4,5, and 6.  These models are based on measurements taken during the full instrument integration and test campaign as described in Sect.~\ref{sec:bandpass}.  The \spherex\ spectral channels are schematically indicated in the leftmost columns of these two figures.
\begin{acknowledgements}
The authors acknowledge many extremely helpful discussions about this software with members of the \spherex\ science and engineering teams.

We acknowledge support from the \spherex\ project under a contract from the NASA/Goddard Space Flight Center to the California Institute of Technology.

Part of the research described in this paper was carried out at the Jet Propulsion Laboratory, California Institute of Technology, under a contract with the National Aeronautics and Space Administration (80NM0018D0004).

The High Performance Computing resources used in this investigation were provided by funding from the JPL Information and Technology Solutions Directorate.

The authors acknowledge the Texas Advanced Computing Center (TACC)\footnote{\url{http://www.tacc.utexas.edu}} at The University of Texas at Austin for providing computational resources that have contributed to the research results reported within this paper.

The PSF and optical distortion models for this paper were generated with FRED by K. Scott Ellis of Photon Engineering.  

This publication makes use of data products from the Wide-field Infrared Survey Explorer, which is a joint project of the University of California, Los Angeles, and the Jet Propulsion Laboratory/California Institute of Technology, and NEOWISE, which is a project of the Jet Propulsion Laboratory/California Institute of Technology. WISE and NEOWISE are funded by the National Aeronautics and Space Administration. 

This publication makes use of data products from the Two Micron All Sky Survey, which is a joint project of the University of Massachusetts and the Infrared Processing and Analysis Center/California Institute of Technology, funded by the National Aeronautics and Space Administration and the National Science Foundation.

This work has made use of data from the European Space Agency (ESA) mission Gaia (https://www.cosmos.esa.int/gaia), processed by the Gaia Data Processing and Analysis Consortium (DPAC, https://www.cosmos.esa.int/web/gaia/dpac/consortium). Funding for the DPAC has been provided by national institutions, in particular the institutions participating in the Gaia Multilateral Agreement.

The Legacy Surveys consist of three individual and complementary projects: the Dark Energy Camera Legacy Survey (DECaLS; Proposal ID No. 014B-0404; PIs: David Schlegel and Arjun Dey), the Beijing-Arizona Sky Survey (BASS; NOAO Prop. ID No.2015A-0801; PIs: Zhou Xu and Xiaohui Fan), and the Mayall z-band Legacy Survey (MzLS; Prop. ID No. 2016A-0453; PI: Arjun Dey). DECaLS, BASS and MzLS together include data obtained, respectively, at the Blanco telescope, Cerro Tololo Inter-American Observatory, NSF’s NOIRLab; the Bok telescope, Steward Observatory, University of Arizona; and the Mayall telescope, Kitt Peak National Observatory, NOIRLab. Pipeline processing and analyses of the data were supported by NOIRLab and the Lawrence Berkeley National Laboratory (LBNL). The Legacy Surveys project is honored to be permitted to conduct astronomical research on Iolkam Du’ag (Kitt Peak), a mountain with particular significance to the Tohono O’odham Nation.

NOIRLab is operated by the Association of Universities for Research in Astronomy (AURA) under a cooperative agreement with the National Science Foundation. LBNL is managed by the Regents of the University of California under contract to the U.S. Department of Energy.

This project used data obtained with the Dark Energy Camera (DECam), which was constructed by the Dark Energy Survey (DES) collaboration. Funding for the DES Projects has been provided by the U.S. Department of Energy, the U.S. National Science Foundation, the Ministry of Science and Education of Spain, the Science and Technology Facilities Council of the United Kingdom, the Higher Education Funding Council for England, the National Center for Supercomputing Applications at the University of Illinois at Urbana-Champaign, the Kavli Institute of Cosmological Physics at the University of Chicago, Center for Cosmology and Astro-Particle Physics at the Ohio State University, the Mitchell Institute for Fundamental Physics and Astronomy at Texas A\&M University, Financiadora de Estudos e Projetos, Fundacao Carlos Chagas Filho de Amparo, Financiadora de Estudos e Projetos, Fundacao Carlos Chagas Filho de Amparo a Pesquisa do Estado do Rio de Janeiro, Conselho Nacional de Desenvolvimento Cientifico e Tecnologico and the Ministerio da Ciencia, Tecnologia e Inovacao, the Deutsche Forschungsgemeinschaft and the Collaborating Institutions in the Dark Energy Survey. The Collaborating Institutions are Argonne National Laboratory, the University of California at Santa Cruz, the University of Cambridge, Centro de Investigaciones Energeticas, Medioambientales y Tecnologicas-Madrid, the University of Chicago, University College London, the DES-Brazil Consortium, the University of Edinburgh, the Eidgenossische Technische Hochschule (ETH) Zurich, Fermi National Accelerator Laboratory, the University of Illinois at Urbana-Champaign, the Institut de Ciencies de l’Espai (IEEC/CSIC), the Institut de Fisica d’Altes Energies, Lawrence Berkeley National Laboratory, the Ludwig Maximilians Universitat Munchen and the associated Excellence Cluster Universe, the University of Michigan, NSF’s NOIRLab, the University of Nottingham, the Ohio State University, the University of Pennsylvania, the University of Portsmouth, SLAC National Accelerator Laboratory, Stanford University, the University of Sussex, and Texas A\&M University.

BASS is a key project of the Telescope Access Program (TAP), which has been funded by the National Astronomical Observatories of China, the Chinese Academy of Sciences (the Strategic Priority Research Program “The Emergence of Cosmological Structures” Grant No. XDB09000000), and the Special Fund for Astronomy from the Ministry of Finance. The BASS is also supported by the External Cooperation Program of Chinese Academy of Sciences (Grant No. 114A11KYSB20160057), and Chinese National Natural Science Foundation (Grant No. 12120101003, No. 11433005).

The Legacy Survey team makes use of data products from the Near-Earth Object Wide-field Infrared Survey Explorer (NEOWISE), which is a project of the Jet Propulsion Laboratory/California Institute of Technology. NEOWISE is funded by the National Aeronautics and Space Administration.

The Legacy Surveys imaging of the DESI footprint is supported by the Director, Office of Science, Office of High Energy Physics of the U.S. Department of Energy under Contract No. DE-AC02-05CH1123, by the National Energy Research Scientific Computing Center, a DOE Office of Science User Facility under the same contract; and by the U.S. National Science Foundation, Division of Astronomical Sciences under Contract No. AST-0950945 to NOAO.

The Pan-STARRS1 Surveys (PS1) and the PS1 public science archive have been made possible through contributions by the Institute for Astronomy, the University of Hawaii, the Pan-STARRS Project Office, the Max-Planck Society and its participating institutes, the Max Planck Institute for Astronomy, Heidelberg and the Max Planck Institute for Extraterrestrial Physics, Garching, The Johns Hopkins University, Durham University, the University of Edinburgh, the Queen's University Belfast, the Harvard-Smithsonian Center for Astrophysics, the Las Cumbres Observatory Global Telescope Network Incorporated, the National Central University of Taiwan, the Space Telescope Science Institute, the National Aeronautics and Space Administration under Grant No. NNX08AR22G issued through the Planetary Science Division of the NASA Science Mission Directorate, the National Science Foundation Grant No. AST-1238877, the University of Maryland, Eotvos Lorand University (ELTE), the Los Alamos National Laboratory, and the Gordon and Betty Moore Foundation.
\end{acknowledgements}

\bibliography{simulator}{}
\bibliographystyle{aasjournalv7}

\end{document}